\documentclass[a4paper,11pt]{article}
\usepackage[textwidth=0.8\paperwidth, textheight=0.8\paperheight]{geometry}

\usepackage{authblk}
\usepackage{float}
\usepackage{graphicx}
\usepackage{hyperref}
\usepackage{amsmath}
\usepackage{amssymb}
\usepackage{microtype}
\usepackage[shortlabels]{enumitem}
\usepackage{cancel}
\usepackage{amsbsy}
\usepackage{xcolor}
\usepackage[capitalise]{cleveref}
\usepackage{breakurl}
\usepackage[normalem]{ulem}
\usepackage{mathtools}
\usepackage{csquotes}
\usepackage{lmodern}
\usepackage{bm}
\usepackage{dsfont}
\usepackage{bbold}
\usepackage{empheq}
\usepackage{caption}
\usepackage{subcaption}
\usepackage{enumitem}
\usepackage{bbm}
\usepackage{comment}
\usepackage[dvipsnames, svgnames]{xcolor}
\usepackage{cite}
\usepackage{eso-pic}

\makeatletter
\DeclareRobustCommand{\rcite}[1]{%
  \rcite@aux#1,\@nil{#1}%
}
\def\rcite@aux#1,#2\@nil#3{%
  \if\relax#2\relax
    Ref.~\cite{#3}%
  \else
    Refs.~\cite{#3}%
  \fi
}
\makeatother

\title{Breaking de Sitter Symmetry with Cosmic Topology: \\  Cosmological Role of Infrared Vacuum Freedom}

\author[1]{Anna Negro\thanks{\href{mailto:anna.negro@case.edu}{anna.negro@case.edu}}\hspace{0.4em}}
\author[1]{Kurt Hinterbichler}
\author[1]{Glenn D. Starkman}
\author[2,1,3]{\protect\linebreak Yashar Akrami}
\author[4,5,6]{Stefano Anselmi}
\author[2]{Javier Carr\'on Duque}
\author[2,7]{Mikel Martin Barandiaran}
\author[8]{Andrius Tamosiunas}

\affil[1]{CERCA/ISO, Department of Physics, Case Western Reserve University,
Cleveland, Ohio 44106, USA}
\affil[2]{Instituto de F\'isica Te\'orica (IFT) UAM-CSIC, C/ Nicol\'as Cabrera 13-15, Campus de Cantoblanco UAM, 28049 Madrid, Spain}
\affil[3]{Astrophysics Group \& Imperial Centre for Inference and Cosmology, Department of Physics, Imperial College London, Blackett Laboratory, Prince Consort Road, London SW7 2AZ, United Kingdom}
\affil[4]{INFN, Sezione di Padova, via Marzolo 8, I-35131 Padova, Italy}
\affil[5]{Dipartimento di Fisica e Astronomia ``G. Galilei'', Universit\`a degli Studi di Padova, via Marzolo 8, I-35131 Padova, Italy}
\affil[6]{Laboratoire Univers et Th\'eories, Observatoire de Paris, Universit\'e PSL, Universit\'e Paris Cit\'e, CNRS, F-92190 Meudon, France}
\affil[7]{Departamento de F\'isica Te\'orica, Universidad Aut\'onoma de Madrid, 28049 Madrid, Spain}
\affil[8]{Institute of Theoretical Astrophysics, P.O. Box 1029 Blindern, N-0315 Oslo, Norway}

\date{}

\makeatletter
\renewcommand\@author{%
  \AB@authlist\\[0.5em]
  \normalfont (COMPACT Collaboration)\\[0.5em]
  \includegraphics[width=0.12\textwidth]{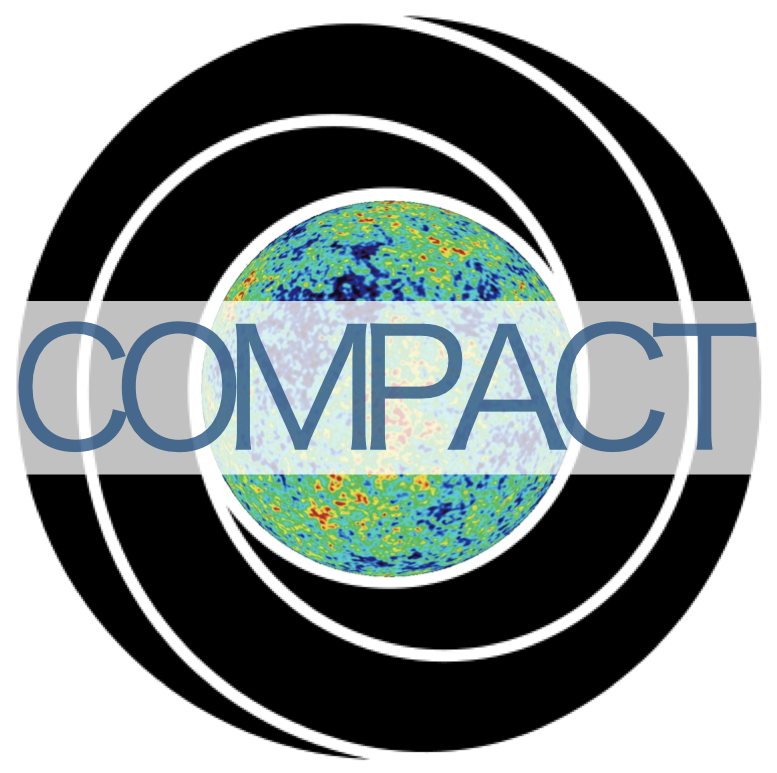}\\[\affilsep]
  \AB@affillist
}
\makeatother

\begin{document}

\maketitle
\vspace{-20pt}

\begin{abstract}

When spatial sections of spacetime are topologically nontrivial, the symmetry arguments usually used to choose the Bunch-Davies vacuum are no longer guaranteed to apply. We study this problem for a scalar field on de Sitter spacetime with toroidal spatial sections. The Hadamard conditions,  which ensure the same short-distance structure as in flat space, are local and naturally insensitive to the compactification. By contrast, we find that the infrared sector is affected by the toroidal identifications that break global de Sitter invariance, thereby removing the usual argument for discarding zero-mode contributions. We treat the remaining zero-mode freedom as an effective topology-induced source and analyze its effects on an inflationary de Sitter expansion within a semiclassical setting. 
We show that, for a massless scalar, the zero-mode sector can contain a contribution to the stress-energy tensor with equation of state $P=-\rho$, behaving as a cosmological constant. Thus, the source driving an early-Universe de Sitter phase need not be determined solely by local terms in the action, but can also depend on the infrared structure of the quantum state. Nontrivial topology, as a concrete realization of de Sitter symmetry violation, turns the otherwise-unfixed infrared freedom in vacuum selection to a contribution directly relevant to the semiclassical expansion. This contribution could potentially be applicable to the current accelerated expansion of the Universe.

\end{abstract}

\tableofcontents

\section{Introduction}

In quantum field theory (QFT) on Minkowski spacetime, the quantum vacuum is selected in a natural way as the lowest-energy state that minimizes the Hamiltonian density. In a generic curved spacetime, the Hamiltonian is time-dependent, with the consequence that ``the lowest energy state" is not univocally defined. As a result, QFT in curved spacetime does not come with a universal prescription that selects a preferred vacuum state \cite{Birrell:1982ix,DeWitt:1975ys}. This freedom, however, is not completely arbitrary. A physically admissible state must ensure the correct short-distance singularity structure to be the same as in flat space. This requirement is encoded in the Hadamard conditions \cite{Hadamard} which fix the ultra-violet (UV) behavior of correlation functions and ensure that local observables can be defined by the usual subtraction of local geometric divergences. (Some examples explicitly using the Hadamard form of the propagator to obtain the renormalized vacuum stress-energy tensor can be found in \rcite{Adler:1976jx,Negro:2024iwy,Breen:2011aa,Tadaki:1987dq,Allen:1987bn,Belokogne:2015etf}.) The Hadamard conditions therefore constrain the local UV structure of the state, but they do not fix the infrared (IR) part of the state. Different Hadamard states may differ by smooth contributions to the two-point function, and these differences can have physical consequences. This IR freedom in selecting a physical quantum vacuum represents one of the intrinsic differences between QFT in Minkowski spacetime and in general curved spacetimes.\\

In exact de Sitter spacetime, the ambiguity in choosing a quantum vacuum is often reduced by imposing additional symmetries or asymptotic conditions \cite{Allen:1987tz,Allen:1985ux,Danielsson:2002qh,Banks:2002nv,Einhorn:2002nu,Kirsten:1993ug}. In the cosmological context, for quantum fields on spatially flat de Sitter backgrounds with simply connected spatial sections, one usually imposes the Bunch-Davies (BD) initial conditions, requiring positive-frequency behavior at short wavelengths in the far past \cite{Bunch:1978yq,Bunch:1977sq}. This prescription naturally selects a Hadamard state and the remaining IR freedom is fixed by requiring the two-point function to be de Sitter invariant, thereby selecting a vacuum that preserves the symmetries of the background. When one discards the zero mode, BD initial conditions select the unique Hadamard, de Sitter-invariant vacuum, which we call the BD vacuum. Therefore, in the simply connected de Sitter case, the spacetime symmetry provides the additional criterion used to remove the residual freedom and single out a unique vacuum to use in inflationary cosmology. Realistic inflationary backgrounds, however, are typically only approximately de Sitter, with the expansion rate slowly evolving in time, so exact de Sitter invariance is not in general available as a criterion for vacuum selection. Beyond spatially flat, exact de Sitter space, different vacuum prescriptions have been studied (e.g., \cite{Chernikov:1968zm,Tagirov:1972vv,Mottola:1984ar,Goldstein:2003ut,Handley,Letey:2022hdp}) motivating a literature on the consequences of non-BD initial conditions as well as on observability of different vacuum choices, with representative examples given in Refs. \cite{Flauger:2013hra,Burgess:2002ub,Bertuzzo:2026gkj,Gasperini:1993yf,Armendariz-Picon:2003knj,Agullo:2012cs,Ganc:2012ae,Meerburg:2009ys,Holman:2007na,Chen:2009bc,Schmidt:2012ky}. More targeted studies have considered settings in which the usual BD prescription is less straightforward, including bouncing inflationary models \cite{Martin:2003sf,Lilley:2011ag,Hwang:2001zt}  and spatially curved universes \cite{Lasenby:2003ur,Letey:2022hdp,Bonga:2016iuf}. \\ 

Here, we turn to a related but less explored setting: how vacuum selection is modified when spacetime is locally de Sitter while its spatial sections have nontrivial global topology \cite{Ford:1989mf,Saharian:2010nep,Lochan:2018pzs,Toussaint:2021czo,Khanna:2011gf}. Any nontrivial flat spatial topology may be obtained by quotienting Euclidean space by a discrete subgroup $\Gamma$ of freely acting elements of the isometry group of $E^3$, the freely acting requirement  ensuring that the quotient $E^3/\Gamma$ is a smooth manifold rather than an orbifold. There are eighteen possible Euclidean spatial topologies that can be found by this quotient construction, labeled as $E_1, \, E_2 \, \cdots \, E_{18}$.
These include compact manifolds ($E_1, \cdots \, E_{10}$), manifolds that are the products of a compact and a non-compact submanifold  ($E_{11}, \cdots \, E_{17}$),  and the simply connected Euclidean space $E_{18}$, which corresponds to the trivial topology usually assumed in cosmology. Most studies of nontrivial cosmic topology have focused on its imprints on the cosmic microwave background and on other signatures arising in the post-inflationary universe, e.g., \rcite{Planck:2015gmu,Cornish:1997ab,Aurich:2013fwa,Bielewicz_2012,Fabre:2013wia,Aurich:2021ofm,Phillips:2004nc,Niarchou:2007nn,COMPACT:2022gbl,Samandar:2025kuf,COMPACT:2024dqe,COMPACT:2024cud,COMPACT:2024qni,COMPACT:2026nez}. The interplay between nontrivial topology and quantum effects in cosmological settings forms a smaller but growing line of work, including recent studies on modifications of power spectra and Casimir backreaction \cite{Fornal:2026qfq,COMPACT:2026pqs,Paraskevas:2025cwu,COMPACT:2026vdj,Godet:2026gqw}. \\

In this work, we ask: what are the consequences of nontrivial spatial topology for the choice of quantum vacuum?  We restrict our attention to $E_1$, the three-torus, as the simplest compact setting in which to study the effect of topology on vacuum selection. This choice keeps the mode structure analytically tractable while retaining the essential features relevant for our question. \\

To address this question, we first revisit the usual vacuum-selection criteria in de Sitter spacetime. We then show how compact spatial topology changes this picture. Since the Hadamard conditions are local, they are insensitive to the global identifications and remain unchanged on the torus. By contrast, the compactification breaks the global de Sitter invariance of the covering spacetime and therefore removes the symmetry argument usually used to discard the zero-mode contribution. The zero mode is then left as a genuine IR ambiguity of the vacuum state. Rather than introducing an additional prescription to fix the zero-mode ambiguity, we treat the zero-mode contribution as an effective topology-induced content of the Universe and study its consequences in the semiclassical Friedmann equation. The goal is therefore not to propose a new vacuum prescription for universes with nontrivial topology. Rather, accepting that the Hadamard conditions leave a set of physically admissible vacua, we take a phenomenological approach and study the consequences of the topology-induced zero-mode freedom. In the semiclassical framework, this freedom enters through the topology-dependent part of the vacuum stress-energy tensor as a source in Einstein's equations. \\

We find that for a massless field on a de Sitter expanding universe the zero-mode sector contains a contribution with equation of state $P=-\rho$, and therefore acts as an effective cosmological constant. In a universe with nontrivial spatial topology, this provides an additional contribution to the vacuum stress-energy tensor whose magnitude is controlled by the IR structure of the quantum state. We study the consequences of this contribution for an early-Universe de Sitter expanding phase within semiclassical gravity. In particular, we show that the topology-dependent vacuum stress-energy tensor can modify an expansion already sourced by a cosmological constant and, for suitable choices of the zero-mode state, can itself support de Sitter expansion in its absence. Reproducing a given inflationary de Sitter  Hubble scale can therefore be reformulated as a condition on the IR sector of the vacuum state. We thereby show how nontrivial topology can turn the otherwise-unfixed IR freedom in vacuum selection into a physically meaningful parameter with potentially observable cosmological consequences.\\

The paper is organized as follows. In \cref{Choosing a vacuum}, we discuss the standard vacuum selection problem for scalar fields in de Sitter spacetime. We first focus on the simply connected covering space, reviewing in \cref{Hadamard conditions trivial topo} the Hadamard conditions and in \cref{BD vacuum trivial topo} the BD vacuum.  We then turn, in \cref{Effects of topo}, to the effects of topology on the choice of a quantum vacuum, considering the three-dimensional torus as a representative example.  In \cref{Independence of the UV}, we verify that compact spatial topology preserves the UV Hadamard structure while in \cref{BD vacuum non trivial topo} we demonstrate that, by breaking the global de Sitter symmetry, nontrivial topology spoils the argument used to remove the zero mode when selecting the BD vacuum. In \cref{effects of zero mode} we treat the zero mode as an effective content of the Universe and study its effects in the framework of semiclassical gravity. We start by discussing the physical interpretation of the zero mode in \cref{Is the zero mode a particle?}; we then show in \cref{topology dep Tmunu} how the scheme-independent, topology-induced contribution to the vacuum stress-energy tensor can be unambiguously extracted; and, in \cref{zero mode EOS}, we examine the zero-mode sector. After studying some special cases, we find that the zero-mode contribution can behave as a cosmological-constant-like fluid. In \cref{semiclassical dS}, we include the full topology-dependent vacuum stress-energy tensor, including this zero-mode contribution, in the semiclassical Friedmann equation and show that reproducing an inflationary Hubble scale can be reformulated as a condition on the IR sector of the vacuum state. We conclude in \cref{Conclusion}.

\section{Choosing a vacuum}\label{Choosing a vacuum}

We start by discussing the choice of quantum vacuum for a scalar field on de Sitter spacetime. We first review the standard vacuum choices adopted in cosmology for the simply connected covering space. We emphasize the distinction between the local Hadamard conditions, which fix the short-distance UV singularity structure needed for the consistent renormalization of local observables, and the additional restriction imposed by requiring de Sitter invariance, which selects the BD vacuum. We then consider the effects of nontrivial topology by focusing on a scalar field on a de Sitter background with compact toroidal spatial sections. Throughout this work, when referring to a de Sitter background or spacetime with nontrivial spatial topology, we mean that the spacetime is locally described by the spatially flat de Sitter metric. The compact spatial identifications, however, break global de Sitter invariance, so that the resulting spacetime does not possess the full de Sitter symmetry of its simply connected covering space. Since the Hadamard conditions are local, whereas topology is an IR property, the local UV singularity structure is not changed by the compactification. The effects of topology appear instead in the IR sector. In particular, the compact spatial identifications break de Sitter invariance globally, preventing the two-point function from retaining the same de Sitter-invariant form of the simply connected covering space. As a consequence, we show that the usual de Sitter-invariance argument used to exclude zero-mode contributions in the simply connected covering space no longer applies.

\subsection{Preliminaries}\label{Preliminaries}

We study a free massive scalar field $\phi$ nonminimally coupled to the background curvature. It is convenient to parametrize the theory in terms of the physical mass $m_{\rm p}$ measured on the exact de Sitter background used in the following calculations (see Appendix \ref{DetailsImprov} for more details),
\begin{equation}\label{actionP}
    S_\phi  = \frac{1}{2}  \int d^4x \sqrt{-g}  \left[ -\nabla_\alpha \phi\nabla^\alpha \phi -m_{\rm p}^2\phi^2 -\xi\left(R-12H^2\right)\phi^2 \right] \, .
\end{equation} 
The background is taken to be a spatially flat de Sitter background 
\begin{equation}
    g_{\alpha \beta}=\operatorname{diag}\left(-a(\eta)^2,a(\eta)^2,a(\eta)^2,a(\eta)^2\right)\,,
\end{equation} 
where the scale factor is $a(\eta)=-1/(H\eta)$, $\eta<0$ denotes conformal time, $R$ is the Ricci scalar, $\xi$ is the non-minimal coupling of the scalar field to the curvature, and $H$ is the de Sitter Hubble parameter. \\

In Fourier space, the equation of motion is
\begin{equation}
    \phi^{\prime \prime} +2\frac{a^{ \prime}}{a} \phi^{ \prime} + |\vec{k}|^2  \phi +m_{\rm p}^2a^2 \phi=0\,, 
\end{equation}
where a prime denotes differentiation with respect to
conformal time. When considering a spatially flat de Sitter background, the equation simplifies to
\begin{equation}
    \phi^{\prime\prime}  -\frac{2}{\eta}\phi^\prime
    +\left[  |\vec{k}|^2  +\frac{m_{\rm p}^2}{H^2\eta^2}
    \right]\phi=0 \, .
\end{equation}
The field operator can be expanded as
\begin{equation}\label{eq:fieldCovering}
    \phi (\eta,\vec{x}) = \int \frac{d^3\vec{k}}{(2 \pi)^{\frac{3}{2}}}  \:\left[ a_{\vec{k}}  e^{i \vec{k}\cdot \vec{x}} \varphi_{\vec{k}}(\eta)+a^\dagger_{\vec{k}} e^{-i \vec{k}\cdot \vec{x}} \varphi_{\vec{k}}^*(\eta) \right],
\end{equation}
where $a_{\vec{k}}$ and $a^\dagger_{\vec{k}}$ are the annihilation and creation operators, satisfying the canonical commutation relations $[a_{\vec{k}},a^\dagger_{\vec{q}}] =  \delta^{(3)}(\vec{k}-\vec{q})$ and $[a_{\vec{k}},a_{\vec{q}}]   = [a^\dagger_{\vec{k}},a^\dagger_{\vec{q}}] = 0$. The mode functions $\varphi_{\vec{k}}(\eta)$ are obtained by solving the equation of motion, yielding\footnote{
     The case $\nu=0$ is special for $\vec{k}=\vec{0}$, but not of interest to us here.}
\begin{equation}\label{eq:ModeFunctionG}
    \begin{aligned}
        &\varphi_{\vec{k}}(\eta) =\frac{H \sqrt{\pi |\eta|^3}}{2} e^{-i\frac{\pi}{2}\left(\nu+\frac12\right)} H^{(1)}_{\nu}(-|\vec{k}| \eta)  \,, \quad \text{for}~\vec{k}\neq \vec{0} \,, \\
        &\varphi_{\vec{0}}(\eta) = H^{1-\nu}b_{\vec{0}} |\eta|^{\frac{3}{2}-\nu} + H^{1+\nu}\frac{c_{\vec{0}}}{2\nu}|\eta|^{\frac{3}{2}+\nu}  \,, \quad \text{for}~\vec{k}= \vec{0} \quad \text{and}\quad \nu\neq0\, .
    \end{aligned}
\end{equation}
Here $H^{(1)}_{\nu}$ is the Hankel function of the first kind, and we have introduced 
\begin{equation}\label{eq:nu}
    \nu =\sqrt{\frac{9}{4}-\frac{m_{\rm p}^2}{H^2}} \,.
\end{equation}
To find the results in Eq. \eqref{eq:ModeFunctionG}, we imposed BD initial conditions as defined in \rcite{Bunch:1978yq,Bunch:1977sq}, which uniquely fix the integration constants of the nonzero modes by requiring positive-frequency plane-wave behavior at early times. By contrast, the integration constants $b_{\vec{0}}$  and $c_{\vec{0}}$, which must satisfy the Wronskian condition
\begin{equation}\label{Wronskian}
    b_{\vec 0}^{*}c_{\vec 0}   -  b_{\vec 0}c_{\vec 0}^{*} = i  
\end{equation}
to ensure the canonical commutation relations, are left unconstrained by the BD initial conditions, since the zero mode is not fixed by the early-time oscillatory prescription. The corresponding vacuum two-point function is then given by 
\begin{equation}\label{full G}
    \begin{aligned}
        G\left(x, x^{\prime}\right)
        =& \frac{H^2 \pi}{4}  \int \frac{d^3\vec{q}}{(2 \pi)^3}  \:  e^{i \vec{q} \cdot\left(\vec{x}-\vec{x}^{\prime}\right)}\sqrt{ |\eta|^3} \sqrt{ |\eta^{\prime }|^3}  H^{(1)}_{\nu}(-|\vec{q}| \eta) H^{(2)}_{\nu}(-|\vec{q}| \eta^{ \prime})\\
        &+ H^{2-2\nu}|b_{\vec{0}}|^2 |\eta^{\prime}|^{\frac{3}{2}-\nu}|\eta|^{\frac{3}{2}-\nu} + H^{2+2\nu}\frac{|c_{\vec{0}}|^2}{4\nu^2}|\eta^{\prime}|^{\frac{3}{2}+\nu}|\eta|^{\frac{3}{2}+\nu}\\
        &+ \frac{H^2}{2 \nu}  \left( |\eta|^{\frac{3}{2}-\nu} |\eta^\prime|^{\frac{3}{2}+\nu}b_{\vec{0}} c_{\vec{0}}^* + |\eta|^{\frac{3}{2}+\nu} |\eta^\prime|^{\frac{3}{2}-\nu} b_{\vec{0}}^*c_{\vec{0}}\right) .
    \end{aligned}
\end{equation}

To better understand the relation between the vacuum choice and the residual freedom associated with the integration constants of the zero mode, $b_{\vec{0}}$ and $c_{\vec{0}}$, we next discuss the conditions that a physical vacuum should satisfy.

\subsubsection{Hadamard conditions -- UV behavior of the vacuum}\label{Hadamard conditions trivial topo}

A Hadamard vacuum $|\Omega\rangle$ is defined by requiring that, for $x^{\prime}$ sufficiently close to $x$, the two-point function can be written in the Hadamard form \cite{Hadamard,Bernard:1986vc,Decanini:2005gt} 
\begin{equation}\label{hadamard}
   G\left(x, x^{\prime}\right) =  \langle \Omega | \phi(x) \phi(x^\prime) | \Omega \rangle \xrightarrow{x \to x^\prime} \frac{1}{8 \pi^2}\left[\frac{U\left(x, x^{\prime}\right)}{\sigma\left(x, x^{\prime}\right)+i\epsilon}+V\left(x, x^{\prime}\right) \ln \left(\sigma\left(x, x^{\prime}\right)+i\epsilon\right)+W\left(x, x^{\prime}\right)\right].
\end{equation}
The geodesic distance $\sigma(x,x^\prime)$ denotes Synge’s world function, i.e. one half of the squared geodesic distance between $x$ and $x^{\prime}$, which vanishes in the coincidence limit $x^\prime \to x$. This is precisely the short-distance limit in which the UV divergences of the two-point function appear. In the same limit, the functions $U(x,x^\prime)$, $V(x,x^\prime)$, and $W(x,x^\prime)$ are smooth and finite. In particular, $U(x,x^\prime)$ and $V(x,x^\prime)$, which are the coefficients of the UV-divergent part, are fixed locally by the geometry and the field equation, whereas the UV-finite part $W(x,x^\prime)$ contains the state-dependent freedom. Choosing a Hadamard vacuum therefore fixes the UV singularity structure of the two-point function to be of the standard Hadamard form. This ensures that the corresponding divergences in the vacuum stress-energy tensor can be subtracted as in flat space by redefining the coupling constants of the gravitational sector, leaving finite, renormalized quantities.\\

It is possible to verify explicitly that the two-point function in Eq. \eqref{full G} indeed has the Hadamard form of Eq. \eqref{hadamard} in the coincidence limit. After performing the angular integration, the two-point function becomes (see Ref. \cite{Bunch:1978yq} for further details)
\begin{equation}\label{Gcs1}
    \begin{aligned}
        G\left(x, x^{\prime}\right)
        =&\frac{H^2 }{8\pi \left|x -x^{\prime}\right|}\int_0^{\infty} dq \: q \sin \!\left(q\left|x -x^{\prime}\right|\right)\sqrt{ |\eta|^3} \sqrt{ |\eta^{\prime }|^3}  H_\nu^{(1)}\left(-q \eta \right) H_\nu^{(2)}\left(-q \eta^{\prime}\right) +G_{\vec{0}}\left(\eta , \eta^{\prime}\right)\\
        =& \frac{H^2 }{16 \pi}\sec (\pi \nu)\left(\frac{1}{4}-\nu^2\right) {}_2F_1\left(\frac{3}{2}+\nu, \frac{3}{2}-\nu ; 2 ; 1+\frac{\Delta u \Delta v}{4 \eta  \eta^{\prime}}\right)+G_{\vec{0}}\left(\eta , \eta^{\prime}\right),
    \end{aligned}
\end{equation}
where we denote $|\vec q\:|=q$, ${}_2F_1$ is the hypergeometric function, and we have defined
\begin{equation}
    \label{eq:lightcone_variables}
    \Delta u=\eta -\eta^{\prime}+|x -x^{\prime}|, 
    \qquad \Delta v=\eta -\eta^{\prime}-|x -x^{\prime}| \,.
\end{equation}
The zero-mode contribution is
\begin{equation}\label{eq:G0}
    \begin{aligned}
    &G_{\vec{0}}\left(\eta , \eta^{\prime}\right) \equiv 
    H^2 |\eta\eta^{\prime}|^{\frac{3}{2}} 
    \left[
    H^{-2\nu}|b_{\vec{0}}|^2 |\eta\eta^{\prime}|^{-\nu} +
    H^{2\nu}\frac{|c_{\vec{0}}|^2}{4\nu^2}|\eta\eta^{\prime}|^{\nu}
    + \frac{1}{2 \nu}  \left( \left|\frac{\eta^\prime}{\eta}\right|^{\nu} b_{\vec{0}} c_{\vec{0}}^* + \left|\frac{\eta}{\eta^\prime}\right|^{\nu} b_{\vec{0}}^*c_{\vec{0}}\right) 
   \right]\,.
    \end{aligned}
\end{equation}
To isolate the divergent terms in the coincidence limit, $\Delta u , \,\Delta v \to 0$ and $\eta  \to \eta^{\prime}$, we use the fact that the hypergeometric function can be expanded as a series in powers of $\Delta u \Delta v$:
\begin{equation}\label{hypergeometric exp}
    \begin{aligned}
    {}_2F_1\left(\frac{3}{2}+\nu, \frac{3}{2}-\nu ; 2 ; 1+\frac{\Delta u \Delta v}{4 \eta  \eta^{\prime}}\right)
    &=\frac{1}{\Gamma\left(\frac{3}{2}+\nu\right) \Gamma\left(\frac{3}{2}-\nu\right)}
    \left\{ - \left(\frac{\Delta u \Delta v}{4 \eta \eta^{\prime}}\right)^{-1}\right. \\
    &\mkern-150mu 
    + \sum_{n=0}^{\infty} \frac{(-1)^n}{n+1} \frac{\Gamma\left(\frac{3}{2}+\nu+n\right) \Gamma\left(\frac{3}{2}-\nu+n\right)}{\Gamma\left(\frac{1}{2}+\nu\right) \Gamma\left(\frac{1}{2}-\nu\right)\Gamma(n+1)\Gamma(n+1)} 
    \left(\frac{\Delta u \Delta v}{4 \eta  \eta^{\prime}}\right)^n \\ 
    &\mkern-100mu \left.\times\left[\ln \left|\frac{\Delta u \Delta v}{4 \eta  \eta^{\prime}}\right|+\psi\left(\frac{3}{2}+\nu+n\right)+\psi\left(\frac{3}{2}-\nu+n\right)-\psi(1+n)-\psi(2+n)\right]\right\},
    \end{aligned}
\end{equation}
where $\psi(z)=\Gamma^\prime(z)/\Gamma(z)$ is the digamma function. Taking the coincidence limit of Eq. \eqref{Gcs1} using the expansion in Eq. \eqref{hypergeometric exp}, we find that the singular terms have precisely the UV behavior required by the Hadamard form in Eq. \eqref{hadamard}. 
Hence, by requiring positive-frequency plane-wave behavior at early times, one naturally selects a Hadamard vacuum.  Crucially, the zero-mode contribution encoded in $G_{\vec{0}}\left(\eta , \eta^{\prime}\right) $  is smooth in the coincidence limit. It therefore contributes only to the finite Hadamard coefficient $W(x,x^\prime)$ and is not constrained by the Hadamard conditions.\\

To make this more explicit, we specialize the results of this section to the conformally invariant case, $m^2_{\rm p}=2 H^2$ and $\xi =\frac{1}{6}$, for which $\nu=\frac{1}{2}$. The mode functions in Eq. \eqref{eq:ModeFunctionG} reduce to
\begin{equation}\label{conformal}
    \begin{aligned}
        &\varphi_{\vec{k}}(\eta)  =\frac{H \eta}{\sqrt{2 |\vec{k}|}} e^{-i|\vec{k}| \eta}\, , \quad \text{for}~\vec{k}\neq \vec{0}, \\
        &\varphi_{\vec{0}}(\eta) = H^{\frac{1}{2}}\,b_{\vec{0}} |\eta| + H^{\frac{3}{2}}c_{\vec{0}}\eta^2 , \quad \text{for}~\vec{k}= \vec{0}\, ,
    \end{aligned}
\end{equation}
and the two-point function in Eq. \eqref{full G} becomes
\begin{equation}\label{propagatorsUV1}
    \begin{aligned}
        G_{\frac{1}{6}}\left(x, x^\prime \right)
        =& \int \frac{d^3 \vec{q}}{(2 \pi)^3}   \frac{ H^2 \eta \eta^\prime}{2q}  e^{i q \Delta \eta-i \vec{q}\cdot \vec{r}}
        +H\left|b_{\vec{0}}\right|^2 \eta \eta^{\prime}
        +H^3\left|c_{\vec{0}}\right|^2 \eta^2 \eta^{\prime 2}-H^2 b_{\vec{0}} c_{\vec{0}}^* \eta \eta^{\prime 2}
        -H^2 b_{\vec{0}}^* c_{\vec{0}} \eta^2 \eta^{\prime},
    \end{aligned}
\end{equation}
where we have defined $\Delta \eta = \eta^\prime -\eta$ and $\vec{r}= \vec{x}^{\, \prime} -\vec{x}$. Performing the integral, or equivalently taking $\nu=\frac{1}{2}$ in Eqs. \eqref{hypergeometric exp} and \eqref{Gcs1}, yields
\begin{equation}\label{propagatorsUV2}
    \begin{aligned}
     G_{\frac{1}{6}}\left(x, x^\prime \right)
     =&- \frac{ 1}{4  \pi^2 } \frac{H^2 \eta \eta^\prime}{\left(\eta -\eta^\prime \right)^2 - | \vec{x} -\vec{x}^{\, \prime} |^2}
     +H\left|b_{\vec{0}}\right|^2 \eta \eta^{\prime}
     +H^3\left|c_{\vec{0}}\right|^2 \eta^2 \eta^{\prime 2}-H^2 b_{\vec{0}} c_{\vec{0}}^* \eta \eta^{\prime 2}
     -H^2 b_{\vec{0}}^* c_{\vec{0}} \eta^2 \eta^{\prime}.
    \end{aligned}
\end{equation}
This shows explicitly that the singular part of the two-point function is the standard Hadamard singularity proportional to $1/\sigma$, while the zero-mode contribution is smooth and finite in the coincidence limit. 

\subsubsection{Bunch-Davies vacuum -- de Sitter invariance and zero-mode contribution}\label{BD vacuum trivial topo}

Since the Hadamard conditions do not uniquely determine the vacuum state---different Hadamard vacua may differ by the smooth finite contribution $W(x,x^\prime)$---we now show how imposing de Sitter invariance fixes the remaining freedom and uniquely selects the quantum vacuum.\\

For a de Sitter-invariant vacuum, the two-point function must depend only on the invariant geodesic separation. Introducing
\begin{equation}\label{eq:Z}
    Z(x, x^\prime) =  1+\frac{\left(\eta-\eta^{\prime}\right)^2-\left|\vec{x}-\vec{x}^{\, \prime}\right|^2}{2 \eta \eta^{\prime}}\, ,
\end{equation}
which is invariant under the de Sitter isometry group $SO(1,4)$, de Sitter invariance requires
\begin{equation}
    G\left(x, x^\prime \right)=G\left(Z(x, x^\prime) \right).
\end{equation}
For a conformally coupled scalar field the two-point function \cref{propagatorsUV2} can be rewritten as
\begin{equation}\label{eq:propagatorCSconfdS}
    \begin{aligned}
        G\left(x, x^{\prime}\right)
        =& \frac{H^2}{8 \pi^2} \frac{1}{1-Z(x, x^\prime)} +H\left|b_{\vec{0}}\right|^2 \eta \eta^{\prime} +H^3\left|c_{\vec{0}}\right|^2 \eta^2 \eta^{\prime 2} -H^2 b_{\vec{0}} c_{\vec{0}}^* \eta \eta^{\prime 2} -H^2 b_{\vec{0}}^* c_{\vec{0}} \eta^2 \eta^{\prime}.
    \end{aligned}
\end{equation}
The first term is explicitly a function of the invariant variable $Z(x, x^\prime)$, whereas the zero-mode contributions depend separately on $\eta$ and $\eta^\prime$ and consequently break the de Sitter invariance.  Therefore, in the simply connected covering space, imposing de Sitter invariance requires discarding the zero-mode contribution. With this choice, the de Sitter-invariant two-point function is recovered:
\begin{equation}
    G\left(x, x^{\prime}\right)= \frac{H^2}{8 \pi^2} \frac{1}{1-Z(x, x^\prime)}\,.
\end{equation}
We conclude that, among the Hadamard states on the simply connected covering space, de Sitter invariance selects the BD two-point function by eliminating the residual zero-mode freedom.\footnote{By expressing $Z(x, x^\prime)$ in terms of $\frac{4 \eta  \eta^{\prime}}{\Delta u \Delta v}$, the same derivation can be extended beyond the conformally coupled case.}

\subsection{Effects of topology in choosing a vacuum}\label{Effects of topo}

Having reviewed the vacuum structure on the simply connected covering space, we now turn to a spatially flat, locally de Sitter universe with compact toroidal spatial sections. Our goal is to understand how nontrivial global topology affects the vacuum choices discussed in \cref{Hadamard conditions trivial topo,BD vacuum trivial topo}. 
We find that the Hadamard conditions are unchanged relative to the covering space: once the modes are required to have positive-frequency plane-wave behavior at early times, the short-distance singular structure remains of the Hadamard form. However, compact topology modifies the IR contribution, and, in particular, we demonstrate that without de Sitter invariance to rely on we have no reason to exclude the zero mode. \\

As for \cref{Hadamard conditions trivial topo,BD vacuum trivial topo}, we consider a massive scalar field on locally pure de Sitter and allow for a nonminimal coupling $\xi$ to the background curvature. For toroidal spatial slices, however, the spatial coordinates are identified under translations by the lattice vectors $\{\vec T_1,\vec T_2,\vec T_3\}$ and the discrete allowed momenta are fixed by the toroidal boundary conditions and satisfy
\begin{equation}\label{eq:kn}
    \vec{k}_{\vec{n}} \cdot \vec{T}_j = 2 \pi n_j\,, \qquad n_j \in \mathbb{Z}, \qquad j=1,2,3.
\end{equation}
The mode equation therefore remains
\begin{equation}
    \phi^{\prime\prime}  -\frac{2}{\eta}\phi^\prime
    +\left[  |\vec{k}_{\vec{n}}|^2 +\frac{m_{\rm p}^2}{H^2\eta^2}   \right]\phi=0 \, ,
\end{equation}
but where $\vec{k}_{\vec n}$ now  labels the discrete set of momentum modes compatible with periodicity on the torus. 
The field operator is expanded as
\begin{equation}\label{eq:FieldTorus}
    \phi (\eta,\vec{x}) = \frac{1}{\sqrt{\mathrm{Vol}}}\sum_{\vec{n}} \left[ a_{\vec{n}}  e^{i \vec{k}_{\vec{n}}\cdot \vec{x}} \varphi_{\vec{n}}(\eta)+a^\dagger_{\vec{n}} e^{-i \vec{k}_{\vec{n}}\cdot \vec{x}} \varphi_{\vec{n}}^*(\eta) \right] \, ,
\end{equation}
where $a_{\vec n}$ and $a^\dagger_{\vec n}$ are the annihilation and creation operators normalized as $[a_{\vec{n}},a^\dagger_{\vec{m}}]= \delta_{\vec{n},\vec{m}}$ 
where the volume of the fundamental domain is $\mathrm{Vol} \equiv \left| \det(\vec T_1,\vec T_2,\vec T_3)\right|$.
The corresponding mode functions are the same as in the simply connected covering space, given in Eq. \eqref{eq:ModeFunctionG}, with the continuum momentum $\vec{k}$ replaced by the reciprocal lattice vectors $\vec{k}_{\vec n}$
\begin{equation}\label{eq:ModeFunctionGTP}
    \begin{aligned}
        &\varphi_{\vec{k}_{\vec n}}(\eta) =\frac{H \sqrt{\pi |\eta|^3}}{2} e^{-i\frac{\pi}{2}\left(\nu+\frac12\right)} H^{(1)}_{\nu}(-|\vec{k}_{\vec n}| \eta)  \,, \quad \text{for}~\vec{k}_{\vec n}\neq \vec{0} \,, \\
        &\varphi_{\vec{0}}(\eta) = H^{1-\nu}b_{\vec{0}} |\eta|^{\frac{3}{2}-\nu} + H^{1+\nu}\frac{c_{\vec{0}}}{2\nu}|\eta|^{\frac{3}{2}+\nu}  \,, \quad \text{for}~\vec{k}_{\vec n}= \vec{0} \, .
    \end{aligned}
\end{equation}

\subsubsection{Independence of the UV from the underlying topology}\label{Independence of the UV}

We demonstrate that the two-point function of a scalar field on a torus can be decomposed into the covering-space result, reviewed in \cref{Preliminaries}, plus an additional contribution due to the nontrivial topology. Importantly, we highlight that the topology correction is UV-finite in the coincidence limit, which is to be expected since the coincidence limit probes arbitrarily small neighborhoods, where the toroidal spacetime is locally identical to its simply connected covering space. \\

Using Eqs. \eqref{eq:FieldTorus} and \eqref{eq:ModeFunctionGTP}, the two-point function is given by 
\begin{equation}\label{eq:G1}
    \begin{aligned}
        G\left( x, x^{\prime }\right)
        =& \frac{H^2 \pi}{4 \mathrm{Vol}}  \sum_{\vec{n}\neq \vec 0}  e^{i \vec{k}_{\vec{n}} \cdot\left(\vec{x}-\vec{x}^{\prime}\right)}\sqrt{ |\eta|^3} \sqrt{ |\eta^{\prime }|^3}  H^{(1)}_{\nu}(-|\vec{k}_{\vec{n}}| \eta) H^{(2)}_{\nu}(-|\vec{k}_{\vec{n}}| \eta^{ \prime})+ \frac{1}{\mathrm{Vol}}\,G_{\vec{0}}\left(\eta, \eta^{\prime}\right)\,,
    \end{aligned}
\end{equation}
where $G_{\vec 0}$ is the zero-mode contribution defined in Eq. \eqref{eq:G0}. To separate the covering-space part from the topology-dependent part, we use the Poisson summation formula \cite{cdadf853-2bd9-3849-9928-547f47a288f0}, which, for the lattice generated by $\{\vec T_1,\vec T_2,\vec T_3\}$, relates the discrete momentum sum to an image sum in position space:
\begin{equation}\label{eq:Poisson}
    \frac{1}{\mathrm{Vol}} \sum_{\vec n} f(\vec k_{\vec n})\,e^{i\vec k_{\vec n}\cdot \vec x} = \sum_{\vec r\in\mathbb Z^3}\int_{\mathbb R^3}\frac{d^3 \vec q}{(2\pi)^3}\,f(\vec q)\,e^{i\vec q\cdot(\vec x+\vec l_{\vec r})}\,,  \qquad   \vec l_{\vec r}=\sum_{j=1}^3 r_j \vec T_j\,.
\end{equation}
We obtain
\begin{equation}\label{eq:G2}
    \begin{aligned}
        G\left(x, x^{\prime }\right)
        =&\frac{H^2 \pi}{4}  \int \frac{d^3\vec{q}}{(2 \pi)^3}  \:  e^{i \vec{q} \cdot\left(\vec{x}-\vec{x}^{\prime}\right)}\sqrt{ |\eta|^3} \sqrt{ |\eta^{\prime }|^3}  H^{(1)}_{\nu}(-|\vec{q}| \eta) H^{(2)}_{\nu}(-|\vec{q}| \eta^{ \prime}) \\
        &+\frac{H^2 \pi}{4} \:\: \widetilde{\sum_{\vec{r}\neq0}} \: \int \frac{d^3 \vec{q}}{(2 \pi)^3} \:  e^{i \vec{q} \cdot\left(\vec{l}_{\vec{r}} +\vec{x}-\vec{x}^{\prime}\right)}\sqrt{ |\eta|^3} \sqrt{ |\eta^{\prime }|^3}  H^{(1)}_{\nu}(-|\vec{q}| \eta) H^{(2)}_{\nu}(-|\vec{q}| \eta^{ \prime}) \\
        &+ \frac{1}{\mathrm{Vol}}\,G_{\vec{0}}\left(\eta, \eta^{ \prime}\right)\,,
    \end{aligned}
\end{equation}
where we introduce the tilded notation which denotes the IR-finite part of the Poisson-resummed expression. This prescription is inherited from the original spectral representation, which is IR finite since the $\vec n=\vec 0$ mode is excluded and treated separately. The apparent IR divergence arising in the Poisson-resummed representation is thus not a divergence of the original nonzero-mode contribution, but results from isolating the continuum/image-space terms without retaining explicitly the subtraction associated with the excluded zero mode. In \cref{zero mode EOS} and Appendix \ref{PoissonEwald}, we give an explicit prescription for extracting the IR-finite part. The first term in Eq. \eqref{eq:G2} is precisely the contribution obtained in Eq. \eqref{Gcs1} for the simply connected covering space. As shown in \cref{Hadamard conditions trivial topo}, its coincidence limit singularity has the Hadamard form. We therefore define the topological correction as the second line of Eq. \eqref{eq:G2}
\begin{equation}\label{Gtp}
    \begin{aligned}
        G^{\mathrm{tp}}\left(x, x^{\prime}\right)
        \equiv
        \frac{H^2 \pi}{4} \: \: \widetilde{\sum_{\vec{r}\neq0}} \: \int \frac{d^3\vec{q}}{(2 \pi)^3} \:  e^{i \vec{q} \cdot\left(\vec{l}_{\vec r}+\vec{x}-\vec{x}^{\prime}\right)}\sqrt{ |\eta|^3} \sqrt{ |\eta^{\prime }|^3}  H^{(1)}_{\nu}(-|\vec{q}| \eta) H^{(2)}_{\nu}(-|\vec{q}| \eta^{ \prime}) 
    \end{aligned}
\end{equation}
and, after integrating over angles, and writing $|\vec q \: |=q$, we obtain
\begin{equation}\label{Gtp1}
    \begin{aligned}
        G^{\mathrm{tp}}\left(x, x^{\prime}\right)
        =&\frac{H^2 }{8\pi \left|\vec l_{\vec r}+ \vec{x}-\vec{x}^{\, \prime}\right|} \widetilde{\sum_{\vec{r}\neq0}} \: \int_0^{\infty} dq \: q \sin \!\left(q\left|\vec l_{\vec r}+ \vec{x}-\vec{x}^{\, \prime}\right|\right)\sqrt{ |\eta|^3} \sqrt{ |\eta^{\prime }|^3}  H_\nu^{(1)}\left(-q \eta\right) H_\nu^{(2)}\left(-q \eta^{\prime}\right)\\
        =& \frac{H^2 }{16 \pi}\sec (\pi \nu)\left(\frac{1}{4}-\nu^2\right)  \widetilde{\sum_{\vec{r}\neq0}} \:  {}_2F_1\left(\frac{3}{2}+\nu, \frac{3}{2}-\nu ; 2 ; 1+\frac{\Delta u_{\vec r} \Delta v_{\vec r}}{4 \eta \eta^{\prime}}\right) \, ,
    \end{aligned}
\end{equation}
where now
\begin{equation}
    \Delta u_{\vec r}=\eta-\eta^{\prime}+|\vec l_{\vec r}+\vec{x}-\vec{x}^{\, \prime}|\,, \qquad \Delta v_{\vec r}=\eta-\eta^{\prime}-|\vec l_{\vec r}+\vec{x}-\vec{x}^{\, \prime}| \, .
\end{equation}
Using the expansion in Eq. \eqref{hypergeometric exp}, we can safely take the coincidence limit. Indeed, for every $\vec r\neq 0$,
\begin{equation}
    \Delta u_{\vec r} \to |\vec l_{\vec r}|\,, \qquad \Delta v_{\vec r} \to -|\vec l_{\vec r}|\,,
\end{equation}
so the argument of the hypergeometric expansion never approaches the singular coincidence-limit point. We therefore find that the UV divergences are entirely encoded in the covering-space contribution, while the topological correction in Eq. \eqref{Gtp} remains finite in the coincidence-limit. This is an example of a general result \cite{Birrell:1982ix}: the short-distance singular structure of the two-point function is insensitive to the underlying toroidal topology. In particular, if one chooses a vacuum whose two-point function has the Hadamard form on the simply connected covering space, then the topological image sum does not spoil this behavior and the corresponding vacuum state remains Hadamard in the toroidal spatial foliation.\\

Returning to the example of a conformally coupled scalar field, the separation between the covering-space contribution, the image contributions, and the zero mode can be seen explicitly. Specializing Eq. \eqref{eq:G2} to $m^2_{\rm p}=2 H^2$ and $\xi=\frac16$, we obtain
\begin{equation}\label{fullpropc}
    \begin{aligned}
        G_{\frac{1}{6}}\left(x, x^{\prime}\right)
        =&- \frac{ 1}{4  \pi^2 } \frac{H^2 \eta \eta^\prime}{\left(\eta-\eta^\prime \right)^2 - | \vec{x} -\vec{x}^{\, \prime} |^2}
        - \frac{H^2 }{4 \pi^2} \: \: \widetilde{\sum_{\vec{r}\neq0}}  \frac{\eta \eta^{\prime}}{\left(\eta-\eta^{\prime}\right)^2-|\vec{l}_{\vec{r}}+\vec{x}-\vec{x}^{\, \prime}|^2} \\
        &+ \frac{1}{\mathrm{Vol}} \left[H|b_{\vec{0}}|^2\eta \eta^\prime +H^3|c_{\vec{0}}|^2\eta^2 \eta^{\prime 2} - H^2 (b_{\vec{0}} c_{\vec{0}}^*)\eta \eta^{\prime 2} - H^2 (b_{\vec{0}}^* c_{\vec{0}})\eta^{2} \eta^{\prime }\right]\,.
    \end{aligned}
\end{equation}
This result makes the structure particularly transparent: the first term is the covering-space propagator and carries the standard Hadamard singularity in the coincidence limit; the second term is the UV-finite image contribution induced by the compact topology; and the last term is the zero-mode contribution, which is smooth and remains unconstrained by the Hadamard conditions. 

\subsubsection{Topology-induced symmetry breaking and zero-mode contributions}\label{BD vacuum non trivial topo}

It is now straightforward to check whether the toroidal propagator can still be written in a de Sitter-invariant form. Rewriting Eq. \eqref{fullpropc} in terms of $Z(x, x^\prime)$ defined in Eq. \eqref{eq:Z}, we obtain 
\begin{equation}\label{fullpropcdS}     
    \begin{aligned} 
        G_{\frac{1}{6}}\left(x, x^{\prime}\right)=&\frac{H^2}{8 \pi^2} \frac{1}{1-Z(x, x^\prime) } - \frac{H^2 }{4 \pi^2} \: \: \widetilde{\sum_{\vec{r}\neq0}} \frac{\eta \eta^{\prime}}{\left(\eta-\eta^{\prime}\right)^2-|\vec{l}_{\vec{r}}+\vec{x}-\vec{x}^{\, \prime}|^2} \\ &+ \frac{1}{\mathrm{Vol}} \left[H|b_{\vec{0}}|^2\eta \eta^\prime +H^3|c_{\vec{0}}|^2\eta^2 \eta^{\prime 2} - H^2 (b_{\vec{0}} c_{\vec{0}}^*)\eta \eta^{\prime 2} - H^2 (b_{\vec{0}}^* c_{\vec{0}})\eta^{ 2} \eta^{\prime }\right] \,.
    \end{aligned} 
\end{equation}
The first term is explicitly de Sitter invariant, as expected for the covering-space contribution to the propagator. By contrast, the second term depends explicitly on the image vectors $\vec{l}_{\vec{r}}$ and therefore cannot be expressed as a function of $Z(x, x^\prime)$ alone. This shows that, not only does the zero-mode contribution provide a source that breaks de Sitter invariance, as for the result in Eq. \eqref{eq:propagatorCSconfdS}, but also  the toroidal identifications break de Sitter invariance at the level of the topological image sum. This is the expected topological breaking of de Sitter invariance due to the compactification. Importantly, even if the zero-mode contribution was discarded, the image term would still remain and de Sitter invariance cannot be restored. We conclude that the two-point function cannot be written solely as a function of the de Sitter-invariant variable $Z(x, x^\prime)$ and that, unlike in the simply connected covering space, de Sitter invariance can no longer be used as a criterion to drop the zero-mode contribution.

\section{Physical effects of the zero mode}\label{effects of zero mode}

Having shown that the usual appeal to de Sitter-invariance to justify discarding the zero-mode contribution is no longer available in a universe with nontrivial spatial topology, we now turn to the physical role of the resulting zero-mode freedom. 
One possible route would be to seek a different prescription that fixes the integration constants $b_{\vec 0}$ and $c_{\vec 0}$, thereby removing the residual zero-mode freedom and selecting a vacuum. Here we take a different approach. Since there is, in general, no canonical choice of vacuum in QFT in curved spacetime, we treat this unfixed zero-mode contribution as an effective phenomenological source and study its consequences in the framework of semiclassical gravity. We compute the vacuum stress-energy tensor of a scalar field on de Sitter background, isolate the scheme-independent, finite, topology-dependent contributions, and determine the effective equation of state associated with different choices of the zero-mode integration constants. We find that the zero-mode sector can generate contributions with $P=-\rho$. Including these together with the remaining topology-dependent contributions to the vacuum stress-energy tensor, we analyze their roles both as corrections to a de Sitter expansion sourced by an existing cosmological constant and as the sole source driving an early-Universe de Sitter expanding phase. We demonstrate that, in universes with nontrivial topology, the zero-mode sector can provide an additional contribution to the cosmological constant whose magnitude is controlled by the IR structure of the quantum vacuum.

\subsection{Quantization and interpretation of the zero mode}\label{Is the zero mode a particle?}

The zero mode is a genuine physical degree of freedom, but in general it does not admit the same interpretation as a harmonic oscillator in field space as the nonzero modes do. For $\vec{k}_{\vec n}\neq\vec{0}$, the modes behave as a harmonic oscillator, oscillating in time with a specific frequency.  This allows us to impose the BD condition on the mode functions, as this is a condition requiring a specific positive frequency at early times.   This in turn leads to the usual oscillator interpretation of $a_{\vec n}$ and $a_{\vec n}^{\dagger}$ as creation and annihilation operators for excitations about the BD vacuum.  By contrast, the massless homogeneous mode has no analogous oscillatory positive-frequency early-time limit and, in the minimally coupled case, is more naturally described as a free-particle degree of freedom in field space, rather than a harmonic oscillator.  A normalizable state of the zero mode may nevertheless be inherited from an earlier regime in which the mode had a nonzero effective mass and an oscillator interpretation was available. The purpose of this section is to clarify these different descriptions and the physical meaning of the zero-mode contribution studied in \cref{zero mode EOS,semiclassical dS}.  \\

Splitting the field into its zero-mode and nonzero-mode contributions, we write
\begin{equation}\label{mode split}
    \begin{aligned}
        \phi (\eta,\vec{x}) &=\phi_{\vec 0} (\eta) + \phi_{\neq \vec 0} (\eta,\vec{x}) \\
        &=   \frac{1}{\sqrt{\mathrm{Vol}}} \left[ a_{\vec{0}} \varphi_{\vec{0}}(\eta)+a^\dagger_{\vec{0}} \varphi_{\vec{0}}^*(\eta) \right] + \frac{1}{\sqrt{\mathrm{Vol}}}\sum_{\vec{n}\neq \vec{0}} \left[ a_{\vec{n}}  e^{i \vec{k}_{\vec{n}}\cdot \vec{x}} \varphi_{\vec{n}}(\eta)+a^\dagger_{\vec{n}} e^{-i \vec{k}_{\vec{n}}\cdot \vec{x}} \varphi_{\vec{n}}^*(\eta) \right]\, ,
    \end{aligned}
\end{equation}
and canonical quantization imposes the Wronskian condition in Eq. \eqref{Wronskian}. For $m_{\rm p}=0$ and $\xi = 0$, and hence $\nu=\frac{3}{2}$, the zero-mode solution in Eq. \eqref{eq:ModeFunctionGTP} results in a linear combination of a constant solution and a solution proportional to $\eta^{3}$.  In this setting, the action is invariant under constant shifts, $\phi\longrightarrow\phi+C $. The homogeneous sector therefore has no potential and is the quantum-mechanical analogue of a free particle rather than a harmonic oscillator. One may still introduce $a_{\vec 0}$ and $a_{\vec 0}^{\dagger}$, as in Eq. \eqref{mode split}, by choosing a complex combination of the two real zero-mode solutions. However, in the absence of an oscillatory positive-frequency condition at early times, no such choice defines a preferred particle basis. In particular, the operator $a_{\vec 0}^{\dagger}a_{\vec 0}$ does not have a preferred interpretation as the number of zero-frequency particles. In this sense, the $a_{\vec 0}$, $a_{\vec 0}^{\dagger}$ parametrization is formally admissible but, as noted in Ref.~\cite{Kirsten:1993ug}, is not the natural description of the zero-mode variables.\\

For a free particle, we could instead introduce the canonical position and momentum variables in field space, $\widehat{x}_0$ and $\widehat{p}_0$. Using the normalization
$[a_{\vec 0},a_{\vec 0}^{\dagger}]=1$ adopted in
Eq. \eqref{mode split}, we can define
\begin{equation}
    \widehat{x}_0   = \frac{1}{\sqrt{H}}\left(   b_{\vec 0}a_{\vec 0} +  b_{\vec 0}^{*}a_{\vec 0}^{\dagger}
    \right)\,,  \qquad   \widehat{p}_0  =  -\sqrt{H}  \left(   c_{\vec 0}a_{\vec 0} +  c_{\vec 0}^{*}a_{\vec 0}^{\dagger} \right)
\end{equation}
so that the homogeneous degree of freedom may be written as
\begin{equation}
    \phi_{\vec 0}(\eta)  =   \frac{1}{\sqrt{\mathrm{Vol}}}   \left(     \widehat{x}_0  + \frac{H^{2}\eta^{3}}{3}\widehat{p}_0  \right)\, ,\qquad  [\widehat{x}_0,\widehat{p}_0]=i \, .
\end{equation}
The constant and $\eta^{3}$ mode functions have vanishing Klein--Gordon norm separately and a nonvanishing mixed Klein--Gordon product, their operator coefficients consequently form the canonical pair $\widehat{x}_0,\widehat{p}_0$. The Wronskian condition in Eq. \eqref{Wronskian} is therefore equivalent to the canonical commutation relation expressed in the complex basis parametrized by $b_{\vec 0}$ and $c_{\vec 0}$. For a noncompact scalar, $\widehat{x}_0$ takes values on the real line, the state satisfying
\begin{equation}
    \widehat{p}_0|\Omega\rangle=0
\end{equation}
is a generalized momentum eigenstate and it is not normalizable. Its wave function is constant in $\widehat{x}_0$, reflecting the fact that all field values related by the shift symmetry are weighted equally. Consequently, the ordinary field two-point function is not well defined in this state. Nevertheless, observables that are invariant under constant shifts, including field derivatives and the stress-energy tensor, remain well defined \cite{Kirsten:1993ug}. (If the scalar field space is instead compact, the zero-mode coordinate is periodic and its conjugate momentum has a discrete spectrum, so that the state with $\widehat{p}_0=0$ is normalizable \cite{Letsios:2026ypo}.)\\

For non-compact scalars, which is the case studied in this work, a particle interpretation may still be available during an earlier regime in which the effective mass is nonzero. In \cite{Ford:1989mf} the authors consider a field whose mass is nonzero initially and tends to zero at later times. In the initial massive regime, the homogeneous mode is an ordinary zero-momentum oscillator, and $a_{\vec 0}$ and $a_{\vec 0}^{\dagger}$ have their standard harmonic oscillator interpretation. At late times, the same mode becomes a zero-frequency mode, for which particle number is no longer meaningfully defined. The Heisenberg operators and the quantum state nevertheless remain well defined, and the late-time zero-mode observables are determined by the initial state and by the evolution of the effective mass. As we discuss in \cref{semiclassical dS}, the semiclassical evolution considered here provides an analogous setting. The topology-dependent backreaction modifies the spacetime curvature and can thereby induce, through the curvature coupling, a nonzero, time-dependent effective mass for the homogeneous mode, before the system approaches the massless zero-frequency regime. The parameters $b_{\vec 0}$ and $c_{\vec 0}$ may therefore be interpreted as parametrizing the state inherited through this evolution, rather than as occupation numbers of zero-frequency particles. We postpone the analysis of the effective mass and its relation to the semiclassical dynamics to \cref{semiclassical dS}.

\subsection{Topology-dependent contributions of the vacuum stress-energy tensor}\label{topology dep Tmunu}

To quantify the effects of the zero-mode contribution during a de Sitter expanding phase, we need to isolate the finite, scheme-independent, topology-dependent part of the vacuum stress-energy tensor of a quantum field in curved spacetime. As shown for the two-point function in \cref{Independence of the UV}, the local UV-divergent contribution associated with the simply connected covering space can be separated unambiguously from the contribution that depends on the underlying topology. The same procedure can be carried out at the level of the stress-energy tensor.\\

The expectation value of the stress-energy tensor of a scalar quantum field in curved spacetime can be obtained from the two-point function as \cite{Birrell:1982ix}
\begin{equation}
    \langle \Omega | T_{\alpha \beta}^{\phi}(x) | \Omega \rangle  =   \lim_{x^{\prime}\to x} \mathcal{D}_{\alpha \beta^\prime } G\left(x, x^{\prime}\right)\,,
\end{equation}
where $\mathcal{D}_{\alpha \beta^\prime }$ is the differential operator that generates the stress-energy tensor of the scalar field under consideration and primed indices refer to the point $x^{\prime}$. Its precise form depends on the mass and curvature coupling; for example, in the minimally coupled massless case, one has
\begin{equation}
    \mathcal{D}_{\alpha \beta^\prime } = \frac{1}{2} \left(\nabla_\alpha \nabla_{\beta^\prime} +\nabla_{\alpha^\prime} \nabla_\beta -g_{\alpha \beta} \nabla^\sigma \nabla_{\sigma^\prime}\right)\,.
\end{equation}
Therefore, by splitting the two-point function into covering-space and topology-dependent contributions as shown in  \cref{Independence of the UV}, one can write
\begin{equation}
    \langle \Omega | T_{\alpha \beta}^{\phi}(x) | \Omega \rangle  =  T^{\mathrm{cs,ren}}_{\alpha \beta}  + T^{\mathrm{tp}}_{\alpha \beta}\,.
\end{equation}
Here $T^{\mathrm{cs,ren}}_{\alpha \beta}$ denotes the renormalized covering-space contribution and is obtained by acting with $\mathcal{D}_{\alpha \beta^\prime }$ on the renormalized covering space propagator. This part contains the usual local, scheme-dependent terms, such as the trace-anomaly contribution for conformally coupled fields \cite{Wald:1978pj,Capper:1974ic,Brown:1976wc,Brown:1977sj,Duff:1993wm}. By contrast, $T^{\mathrm{tp}}_{\alpha \beta}$ is the finite, scheme-independent, potentially observable contribution which can be obtained by acting with $\mathcal{D}_{\alpha \beta^\prime }$ on the topology-dependent part of the propagator. \\

In the following sections, however, we use an equivalent but more direct route to compute $T^{\mathrm{tp}}_{\alpha \beta}$. Rather than first isolating the topology-dependent part of the propagator and then applying the point-splitting operator, we compute the stress-energy tensor directly from the momentum-space mode expansion of the field. The topology-dependent contribution is then isolated by following the same steps as in \cref{Independence of the UV}. The discrete momentum sums are rewritten in position space using the Poisson summation formula in Eq. \eqref{eq:Poisson}; the UV-divergent $\vec r=\vec 0$ term is discarded, while the remaining scheme-independent $\vec r\neq\vec 0$ image terms yield the finite topology-dependent contribution. We note that in following this route we take the coincidence limit from the outset. This should not be understood as a renormalization of the full vacuum stress-energy tensor, since the covering-space contribution remains UV divergent and must be treated using the usual regularization and renormalization procedure. The shortcut is nevertheless valid for extracting the topology-dependent part: after Poisson resummation, all UV sensitivity resides in the $\vec r=\vec 0$ term, whereas the $\vec r\neq\vec 0$ image terms are finite in the coincidence limit. Discarding the former therefore isolates the finite topology-dependent contribution unambiguously, even though the full coincident expression is not itself a renormalized quantity.  As for the two-point function, this procedure unambiguously distinguishes the two contributions: $T^{\mathrm{cs}}_{\alpha \beta}$, which is the local term determined by the covering-space geometry, and $T^{\mathrm{tp}}_{\alpha \beta}$ which is the finite, scheme-independent term associated with the compact spatial topology. Once the covering-space contribution is removed, the remaining topology-dependent stress tensor is therefore a vacuum, Casimir-type contribution. The zero mode, which is not captured by the nonzero image sum, is treated separately as a finite IR contribution and, since our focus is the zero mode, we further decompose $T^{\mathrm{tp}}_{\alpha \beta}$, by isolating the zero-mode dependent contribution from the image sum
\begin{equation}
    T^{\mathrm{tp}}_{\alpha \beta}  = T^{\mathrm{tp},\vec n\neq \vec 0}_{\alpha \beta}  + T^{\mathrm{tp},\vec 0}_{\alpha \beta}\,.
\end{equation}

\subsection{Equation of state of the zero mode contribution}\label{zero mode EOS}

We now apply the procedure described in Sec.  \ref{topology dep Tmunu} to the vacuum stress-energy tensor of a scalar field on de Sitter background and, in extracting the part induced by the compact spatial topology, we aim to determine the equation of state of the topology-dependent contribution. Particular attention is paid to the zero mode, whose contribution is not constrained by the Hadamard conditions. \\

We begin with two special cases in which the equation of state of the zero-mode contribution can be read off directly: a conformally coupled and a minimally coupled massless scalar field on a locally de Sitter background with cubic-torus spatial sections, $\vec{k}_{\vec n}=\frac{2 \pi}{L}  \vec n $. For the conformally coupled case, the topology-dependent part of the stress-energy tensor is
\begin{equation}\label{eq:ZeroContsC}
    T^{\mathrm{tp},\vec n\neq \vec 0}_{\nu=\frac{1}{2}}{}_{\alpha}{}^{\beta}= -\frac{H^4 \eta^4}{2\pi^2 L^4 }  \sum_{\vec{r}  \in \mathbb{Z}^3 \setminus \{\vec{0}\}} \frac{1}{ |\vec{r}\:|^4}\operatorname{diag}\left(-1, \frac{1}{3},\frac{1}{3},\frac{1}{3}  \right) \, ,\qquad T^{\mathrm{tp},\vec{0}}_{\nu=\frac{1}{2}}{}_{\alpha}{}^{\beta} =  \frac{H^5  \eta^4 |c_{\vec{0}}|^2 }{2 L^3}  \operatorname{diag}\left(-1, \frac{1}{3},\frac{1}{3},\frac{1}{3}  \right)\,.
\end{equation}
We therefore find that, independently of the choice of integration constants that specify the zero mode, the zero-mode contribution is radiation-like, with
\begin{equation}
    P=\frac{1}{3} \rho \, ,
\end{equation}
which is the same equation of state as the nonzero-mode contribution. This, however, is not true in general. As shown in Ref. \cite{COMPACT:2026vdj}, for a rectangular or tilted torus the contribution to the pressure from the nonzero modes can be anisotropic, and the underlying topology can source anisotropic expansion.\\

For a massless minimally coupled scalar field, the zero-mode contribution instead becomes 
\begin{equation}\label{eq:ZeroConts}
    T^{\mathrm{tp},\vec{0}}_{\nu=\frac{3}{2}}{}_{\alpha}{}^{\beta} =  \frac{H^7  \eta^6 |c_{\vec{0}}|^2 }{2 L^3}  \operatorname{diag}\left(-1, 1,1,1  \right) \, .
\end{equation}
In this case the zero mode behaves as a stiff fluid, with equation of state
\begin{equation}
    P=\rho \, .
\end{equation}
We therefore obtain that in both Eq. \eqref{eq:ZeroContsC} and Eq. \eqref{eq:ZeroConts} the integration constants of the zero mode affect the overall magnitude of the topology-induced zero-mode contribution. In the general case, however, the zero mode can generate qualitatively different effective fluids depending on the mass, the curvature coupling, and the choice of integration constants. \\

We next consider the more general case of a massive scalar field with nonminimal coupling $\xi$ to the background curvature. We therefore write
\begin{equation}\label{action}
    \begin{aligned}
        S&= S_{\rm bg}+S_\phi \\
        &=\frac{1}{16\pi G}
        \int d^4x\sqrt{-g}\,
        \left(R-2\Lambda_{\rm eff}\right) +\frac{1}{2}\int d^4x\sqrt{-g} \left[ -\nabla_\sigma\phi\nabla^\sigma\phi -m_{\rm p}^2\phi^2 -\xi\left(R-12H^2\right)\phi^2 \right] \, ,
    \end{aligned}
\end{equation}
where the constant Hubble scale $H$ is the Hubble scale of the reference de Sitter geometry on which the field modes and the vacuum stress-energy tensor are evaluated, and is held fixed when varying the action with respect to the metric. Although the shifted curvature term does not appear in the equation of motion, it still contributes to the improved stress-energy tensor (see Appendix \ref{DetailsImprov} for more details on the improved vacuum stress-energy tensor and the comparison with the standard parametrization). Varying the action in Eq. \eqref{action} with respect to the background metric, we obtain the improved stress-energy tensor 
\begin{equation}
    \begin{aligned}\label{TmunuImprps}
        \langle \Omega | T_{\alpha\beta}^{\rm imp}| \Omega \rangle=\langle \Omega | \nabla_{\alpha}\phi \nabla_{\beta}\phi -\frac{1}{2}g_{\alpha \beta} \nabla_{\sigma}\phi \nabla^{\sigma}\phi -\frac{1}{2}m_{\rm p}^2 g_{\alpha \beta}\phi^2 | \Omega \rangle - \xi \langle \Omega | \Delta T_{\alpha\beta}| \Omega \rangle \, ,
    \end{aligned}
\end{equation}
where
\begin{equation}\label{ImprovementMain}
    \langle \Omega | \Delta T_{\alpha \beta}| \Omega \rangle = \langle \Omega |  \left(   \nabla_\alpha\nabla_\beta  -g_{\alpha \beta}\nabla_{\sigma}  \nabla^{\sigma}  -3H^2g_{\alpha \beta}\right)\phi^2  | \Omega \rangle\, .
\end{equation}
The full stress-energy tensor in Eq. \eqref{TmunuImprps} is covariantly conserved on shell, while the $\xi$-dependent improvement contribution in Eq. \eqref{ImprovementMain} is separately conserved on the exact de Sitter background. Moreover, on the torus, the improvement contribution does not modify the charges associated with the surviving spatial translations, because its contribution to each charge is a total spatial derivative and integrates to zero over the compact spatial slice.  \\

Using the field operator expansion in Eq. \eqref{eq:FieldTorus} and the zero-mode solution in Eq. \eqref{eq:ModeFunctionGTP} we compute the zero-mode contribution to the improved stress-energy tensor in Eq. \eqref{TmunuImprps} for generic mass and curvature coupling, obtaining
\begin{equation}\label{eq:ImprovedZeroModeGeneral}
    \begin{aligned}
        \rho^{\mathrm{imp,tp},\vec{0}}_{\nu} ={}& \frac{H^{4-2\nu}|\eta|^{3-2\nu}}{2 \mathrm{Vol}}  \left[\left(\frac{3}{2}-\nu\right)^2 +\frac{m_{\rm p}^2}{H^2} +12(\nu-2)\xi  \right]  |b_{\vec 0}|^2 \\
        &+ \frac{H^{4+2\nu}|\eta|^{3+2\nu}}{8\nu^2 \mathrm{Vol}} \left[ \left(\frac{3}{2}+\nu\right)^2 +\frac{m_{\rm p}^2}{H^2} -12(2+\nu)\xi \right] |c_{\vec 0}|^2 \\
        &- \frac{H^{2}\eta^{3}}{2\nu \mathrm{Vol}} \left(m_{\rm p}^2-12H^2\xi\right) \left( b_{\vec 0}^*c_{\vec 0} + b_{\vec 0}c_{\vec 0}^* \right)\, , \\[1ex]
        P^{\mathrm{imp,tp},\vec{0}}_{\nu}{}_{i}{}^{j} ={}& \frac{H^{4-2\nu}|\eta|^{3-2\nu}}{2\mathrm{Vol}} \delta_i{}^j \left[ (1-4\xi) \left(  \left(\frac{3}{2}-\nu\right)^2 -\frac{m_{\rm p}^2}{H^2} \right) -4\xi\left(\frac{3}{2}-\nu\right) +6\xi \right] |b_{\vec 0}|^2 \\
        &+ \frac{H^{4+2\nu}|\eta|^{3+2\nu}}{8\nu^2\mathrm{Vol}} \delta_i{}^j \left[ (1-4\xi) \left(  \left(\frac{3}{2}+\nu\right)^2 -\frac{m_{\rm p}^2}{H^2}  \right) -4\xi\left(\frac{3}{2}+\nu\right)+6\xi \right]
        |c_{\vec 0}|^2 \, .
    \end{aligned}
\end{equation}
The zero-mode solution in the massless case, $m_{\rm p} = 0$, $\nu = \frac{3}{2}$, is 
\begin{equation}
    \varphi_{\vec 0}(\eta) = \frac{b_{\vec 0}}{\sqrt{H}} + \frac{\sqrt{H^5}}{3}c_{\vec 0}|\eta|^3 \, ,
\end{equation}
while the stress-energy tensor still retains its dependence on the coupling to the curvature $\xi$ and Eq. \eqref{eq:ImprovedZeroModeGeneral} reduces to
\begin{equation}\label{eq:dm3}
    \begin{aligned}
        \rho^{\mathrm{imp,tp},\vec{0}}_{\nu=\frac{3}{2}} ={}& -\frac{3H\xi}{\mathrm{Vol}}|b_{\vec 0}|^2 +(9-42\xi)\frac{H^7\eta^6}{18\mathrm{Vol}}|c_{\vec 0}|^2 +\frac{4H^4\xi\eta^3}{\mathrm{Vol}} \left(   b_{\vec 0}^*c_{\vec 0}+b_{\vec 0}c_{\vec 0}^* \right),\\
        P^{\mathrm{imp,tp},\vec{0}}_{\nu=\frac{3}{2}}{}_i{}^j ={}&\frac{3H\xi}{\mathrm{Vol}}|b_{\vec 0}|^2\delta_i{}^j+(9-42\xi)\frac{H^7\eta^6}{18\mathrm{Vol}}|c_{\vec 0}|^2\delta_i{}^j \, .
    \end{aligned}
\end{equation}
Hence, we find that the constant zero-mode branch proportional to $b_{\vec 0}$ behaves as a cosmological-constant-like contribution, $P=-\rho$, while the branch proportional to $c_{\vec 0}$ gives $P=\rho$. For generic choices of $b_{\vec 0}$ and $c_{\vec 0}$, the interference term contributes to $\rho$ but not to $P_i{}^j$ and therefore contributes as dust. This result is closely related to the recent analysis of Ref.~\cite{Jimenez-Aguilar:2026aci}, which found no accelerated expansion from scalar-field fluctuations in the simply connected spatially flat FLRW, while identifying negative-pressure contributions in the closed, positively curved case. Our result is complementary: the spatial sections considered here are locally flat, but the toroidal identifications make the zero mode a genuine global degree of freedom. As a result, a cosmological-constant-like sector can arise even without positive spatial curvature, through nontrivial topology.\\

In order to quantify the effects of the zero-mode contribution in Eq. \eqref{eq:dm3} by including it as an effective source of the semiclassical Einstein equations, we compute the total vacuum stress-energy tensor given in Eq. \eqref{TmunuImprps} in the massless limit. We find that the components of $\langle \Omega | T_{\alpha\beta}^{\rm imp}| \Omega \rangle$ are (see Appendix \ref{NonZero} for more details)
\begin{equation}\label{eq:Components1}
    \begin{aligned}
        \rho^{\mathrm{imp}}_{\nu=\frac{3}{2}} = & \frac{H^4}{4 \mathrm{Vol}}\sum_{\vec{n} \neq 0} \left(2 \eta^4 |\vec{k}_n| +\frac{\eta^2}{ |\vec{k}_n|} \right)- \frac{3 H^4\xi}{2 \mathrm{Vol}}\sum_{\vec{n} \neq 0} \left( 3 \frac{\eta^2}{ |\vec{k}_n|} +\frac{1}{ |\vec{k}_n|^3} \right) \\
        & -\frac{3H\xi}{\mathrm{Vol}}|b_{\vec 0}|^2 +(9-42\xi)\frac{H^7\eta^6}{18 \mathrm{Vol}}|c_{\vec 0}|^2 +\frac{4H^4\xi\eta^3}{\mathrm{Vol}} \left(   b_{\vec 0}^*c_{\vec 0}+b_{\vec 0}c_{\vec 0}^* \right) \, ,\\
        P^{\mathrm{imp}}_{\nu=\frac{3}{2}}{}_i{}^j = & \frac{H^4 }{2\mathrm{Vol}  } \sum_{\vec{n} \neq 0} (\vec{k}_n)_i(\vec{k}_n)^j\left(\frac{\eta^4}{ |\vec{k}_n|}+ \frac{\eta^2}{ |\vec{k}_n|^3} \right) -\frac{H^4}{4 \mathrm{Vol}  }\sum_{\vec{n} \neq 0}\frac{\eta^2}{ |\vec{k}_n|}    \delta_i{}^j + \frac{3 H^4\xi}{2\mathrm{Vol}} \sum_{\vec{n} \neq 0} \left( \frac{\eta^2}{ |\vec{k}_n|} +\frac{1}{ |\vec{k}_n|^3} \right) \delta_i{}^j \\
        &+\frac{3H\xi}{\mathrm{Vol}}|b_{\vec 0}|^2\delta_i{}^j+(9-42\xi)\frac{H^7\eta^6}{18\mathrm{Vol}}|c_{\vec 0}|^2\delta_i{}^j \, .
    \end{aligned}
\end{equation}
We next isolate the finite, topology dependent contribution of the nonzero modes. For this purpose, we specialize to a cubic torus for which $\vec{k}= \frac{2 \pi}{L} \vec{n}$, and $\mathrm{Vol}=L^3$. This assumption simplifies the lattice sums but does not qualitatively affect the equation of state of the zero-mode contribution, since the zero mode is spatially homogeneous and depends on the compactification only through the overall factor $1/\mathrm{Vol}$. The situation is different for the nonzero modes: for a rectangular and/or tilted torus, the discrete spectrum depends on the shape of the fundamental domain and can lead to anisotropic components of the stress-energy tensor. These anisotropic effects have been presented in Ref. \cite{COMPACT:2026vdj}.\\

To go from the discrete momentum-space expression in Eq. \eqref{eq:Components1} to the finite topology-dependent result, we use the Poisson summation formula and an Ewald summation technique \cite{Ewald,NIJBOER1957309}. The Poisson summation formula introduced in Eq. \eqref{eq:Poisson} rewrites the discrete momentum sums as sums over spatial images, while the Ewald summation technique separates the result into rapidly convergent real-space and reciprocal-space pieces. This procedure provides an explicit representation for the topology-dependent contributions of the power-divergent sums, while the topology contribution of the  constant contribution is kept formally defined using the general covering-space/topology separation. The details of this derivation, together with the intermediate expressions, are presented in Appendix \ref{PoissonEwald}. We obtain 
\begin{equation}\label{Tmunu TP}
    \begin{aligned}
         \rho^{\mathrm{imp,tp}}_{\nu=\frac{3}{2}} = & - \frac{ \eta^4 H^4}{ 2\pi^2 L^4 }    \sum_{\vec r\neq \vec 0} \frac{1}{|\vec r|^4}  + \left( 1 - 18 \xi \right) \frac{\eta^2 H^4 B_1}{8 \pi  L^2}  - \frac{3 H^4\xi}{16 \pi^3 }  \overbracket[0.8pt][0pt]{\sum_{\vec{n} \neq 0}} \frac{1}{|\vec n|^3} -3H^4 \xi|\tilde{b}_{\vec 0}|^2 +(9-42\xi)\frac{H^4 \eta^6}{18 L^6}|\tilde{c}_{\vec 0}|^2 \, \\
         &+\frac{4H^4\xi\eta^3}{L^3} \left(   \tilde{b}_{\vec 0}^*\tilde{c}_{\vec 0}+\tilde{c}^*_{\vec 0}\tilde{b}_{\vec 0} \right)\,,\\
         P^{\mathrm{imp,tp}}_{\nu=\frac{3}{2}}  =&  - \frac{ \eta^4 H^4}{6\pi^2 L^4} \sum_{\vec r\neq \vec 0}  \frac{1}{|\vec r|^4} - \left(1 -18\xi\right) \frac{\eta^2 H^4 B_1}{24\pi L^2} + \frac{3 H^4\xi}{16 \pi^3 }  \overbracket[0.8pt][0pt]{\sum_{\vec{n} \neq 0}} \frac{1}{|\vec n|^3}  + 3H^4\xi|\tilde{b}_{\vec 0}|^2 + (9-42\xi) \frac{H^4\eta^6}{18L^6} |\tilde{c}_{\vec 0}|^2 \, ,
    \end{aligned}
\end{equation}
where we have rescaled the zero-mode integration constants according to
\begin{equation}\label{rescaling}
b_{\vec 0}=(HL)^{3/2}\tilde b_{\vec 0}\, ,
\qquad
c_{\vec 0}=(HL)^{-3/2}\tilde c_{\vec 0} \, ,
\end{equation}
which preserves the Wronskian condition in Eq. \eqref{Wronskian}. We stress that this rescaling only absorbs the explicit dependence on the compactification scale into the dimensionless zero-mode parameters, however, $b_{\vec 0}$ may itself depend on $L$. We have also introduced the topology-dependent finite quantity $B_1$ defined in Eq. \eqref{brakets} which does not depend on the dimensionless parameter $\lambda L^2$ (see Appendix \ref{Ewald} for more details) 
\begin{equation}\label{brakets}
    \begin{aligned}
        B_1 \equiv {} & \frac{1}{\pi} \sum_{\vec{r} \neq 0} \frac{e^{-\lambda L^2 |\vec{r}|^2}}{|\vec{r}|^2}+\sum_{\vec{n} \neq 0} \frac{\operatorname{erfc}\left(\frac{\pi \left|\vec{n}\right|}{L \sqrt{\lambda}}\right)}{|\vec{n}|} -\frac{2 \sqrt{\pi}}{L } \frac{1}{\sqrt{\lambda}}  - \frac{\lambda L^2}{\pi}  \, ,
    \end{aligned}
\end{equation} 
and introduced the overbarred notation to denote the UV-finite topology-dependent contribution associated with the nonzero mode contribution to the constant term. Finally, since the topology-dependent stress-energy tensor is spatially isotropic for the cubic torus, we have identified the pressure with the trace of its spatial components 
\begin{equation}\label{pressureRed}
    P^{\mathrm{imp,tp}}_{\nu=\frac{3}{2}} \equiv \frac{1}{3} P^{\mathrm{imp,tp}}_{\nu=\frac{3}{2}}{}_i{}^i \, .
\end{equation} 
We emphasize that the derivation of Eq. \eqref{Tmunu TP} follows from the general argument of Sec. \ref{topology dep Tmunu}: the UV-divergent, scheme-dependent contribution is local and therefore belongs to the covering-space part, while the dependence on the nontrivial spatial topology is finite and scheme independent. For the terms that admit the Poisson/Ewald representation, this separation is explicit---the $\vec r = \vec 0$ term gives the covering-space contribution, whereas the $\vec r\neq \vec 0$ terms yield the finite topology-dependent part. For the logarithmically divergent constant term, the same procedure cannot be used to extract the finite contribution \cite{NIJBOER1957309}; we therefore retain the topology-dependent part formally through the overbar notation introduced above. Since its effect is degenerate with the zero-mode contribution, for the purposes of the present work it is sufficient to keep this contribution formally defined. Our aim is not to determine the absolute value of the topology-induced constant vacuum contribution, but to study whether and how the remaining freedom in the IR sector can affect the semiclassical de Sitter expansion. \\

We now have the necessary ingredients to quantify the effects of the presented freedom in the choice of vacuum in a universe with nontrivial spatial topology. In what follows, we include the scheme-independent, topology-induced stress-energy tensor in Eq. \eqref{Tmunu TP} in the semiclassical Einstein equations and study its effect during a de Sitter expanding phase. 

\subsection{Semiclassical de Sitter expansion in a cubic toroidal universe}\label{semiclassical dS}

In this section, we begin by estimating the relative magnitude of the different topology-dependent contributions to the vacuum stress-energy tensor. By identifying which terms dominate at different stages of the de Sitter expanding phase, we are led to a simpler limit in which the energy density is effectively constant and can be compared directly with a vacuum-energy contribution. We then use this limit to distinguish two cases: one in which topology corrects an existing cosmological constant which is sourcing an early Universe de Sitter expansion, and one in which the topology-dependent vacuum stress-energy tensor itself is the only source of the expansion. This setup is not intended as a realistic model of inflation, since an exact de Sitter phase persists indefinitely and contains no mechanism for transitioning to the subsequent cosmological evolution. However, exact de Sitter provides a setting in which we can analytically estimate the effects of the zero mode and the size of the zero-mode amplitude required to reproduce, by itself, a GUT-scale inflationary Hubble rate.\\

\begin{figure}[h]
    \centering
    \includegraphics[width=1\linewidth]{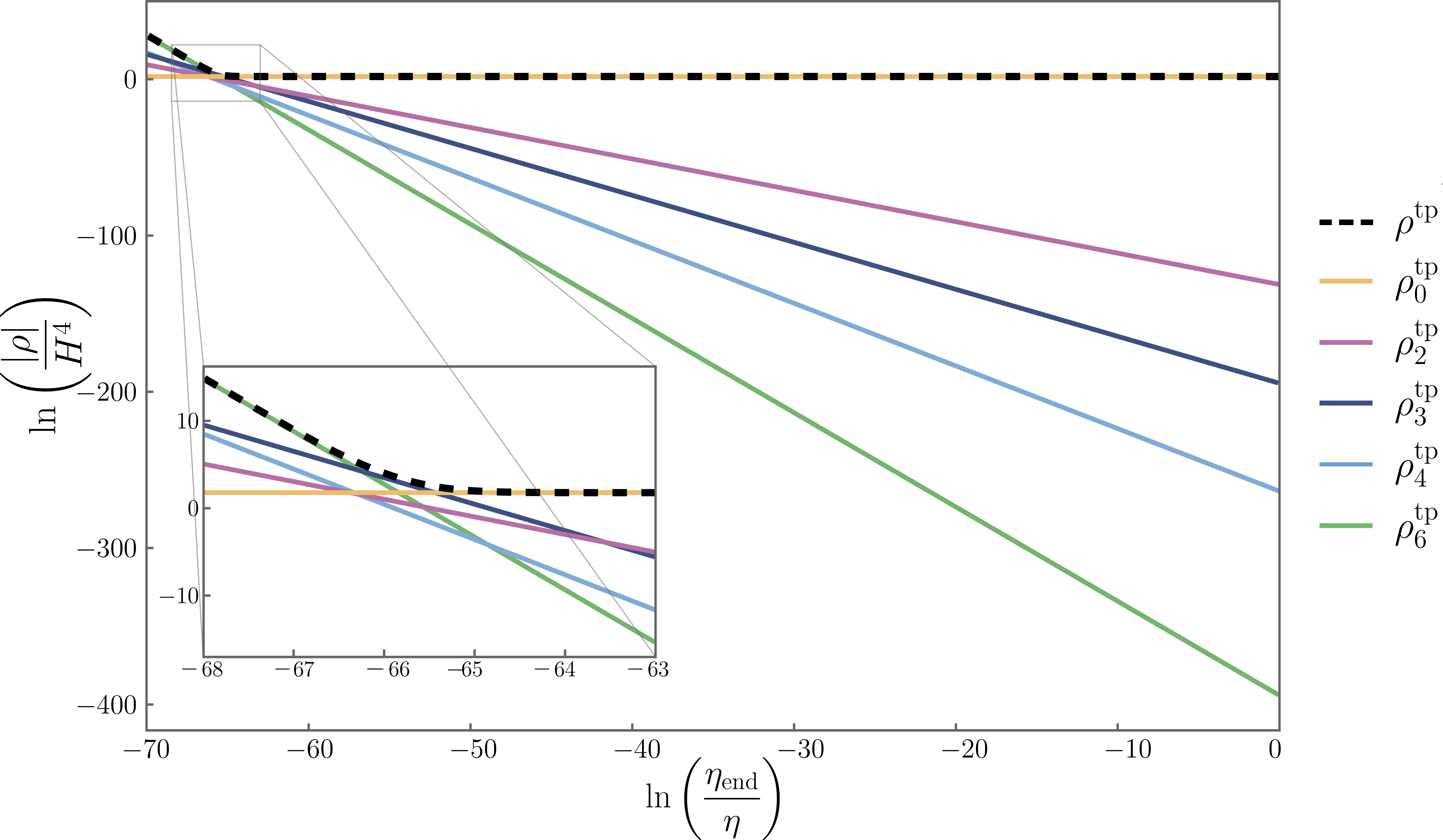}
  \caption{Topology-dependent energy density during an inflationary de Sitter expansion. We plot $\ln\!\left(|\rho|/H^4\right)$ for the total topology-dependent energy density $\rho^{\mathrm{tp}}$, given in Eq. \eqref{Tmunu TP}, together with the different contributions grouped according to their scaling with conformal time, namely $\eta^6$, $\eta^4$, $\eta^3$, $\eta^2$, and $\eta^0$ sectors, as functions of $\ln\!\left(\eta_{\mathrm{end}}/\eta\right)$. We use $H=10^{14}\,\mathrm{GeV}$, $L=4\cdot10^{42}\,\mathrm{GeV}^{-1}$ and take the overbarred contribution, $\xi$, $\tilde b_{\vec 0}$, and $\tilde c_{\vec 0}$ to be of order unity. Numerically evaluating the sum in Eq. \eqref{brakets} we obtain $B_1\simeq-3$. The inset magnifies the early-time region in which the different contributions become comparable. The fact that several curves nearly intersect in this region is not a generic feature, but depends on the choice of the zero-mode integration constants and on the cosmological parameters. Over most of the inflationary interval, the total energy density is dominated by the constant contribution $\rho^{\mathrm{tp}}_0$. This qualitative behavior is robust under increasing $|\tilde b_{\vec 0}|$.}
    \label{fig:plotLog}
\end{figure}

Using that $R_{\alpha\beta}=3 H^2 g_{\alpha\beta}$ and $R=12 H^2$ for flat de Sitter space, the first Friedmann equation obtained by varying the action in Eq. \eqref{action} with respect to the background metric, is
\begin{equation}
    3 H^2 - \Lambda_{\rm eff} = 8 \pi G \rho^{\mathrm{tp}} 
\end{equation}
where, to simplify the notation in this Section, we define $\rho^{\mathrm{tp}} \equiv \rho^{\mathrm{imp,tp}}_{\nu=\frac{3}{2}}$. All purely local/covering-space renormalized contributions are regarded as part of the renormalized couplings, and here we study the finite change induced by topology. The value of the de Sitter scale $H$ introduced in Eq. \eqref{action} is now determined self-consistently by the semiclassical Einstein equation. Fig. \ref{fig:plotLog} shows, on a logarithmic scale, the topology-dependent energy density $\rho^{\mathrm{tp}}$ given in Eq. \eqref{Tmunu TP} and its individual contributions $\rho^{\mathrm{tp}}_n$, with $\rho^{\mathrm{tp}}_n$ denoting the term proportional to $\eta^n$. The corresponding equations of state, $w_n \equiv P^{\mathrm{tp}}_n/\rho^{\mathrm{tp}}_n$, and the origin of each contribution are summarized in Table \ref{tab:topology_contributions}.

\begin{table}[H]
\centering
\renewcommand{\arraystretch}{1.3}
\begin{tabular}{c c c c c}
\hline
\hline
\textbf{Contribution} & \textbf{Scaling} & $\boldsymbol{w_n}$ &
\textbf{Zero-mode dependence} & \textbf{Nonzero modes} \\
\hline
$\rho^{\mathrm{tp}}_6$
& $\eta^6$
& $1$
& $|\tilde c_{\vec 0}|^2$
& --- \\

$\rho^{\mathrm{tp}}_4$
& $\eta^4$
& $1/3$
& ---
& $\checkmark$ \\

$\rho^{\mathrm{tp}}_3$
& $\eta^3$
& $0$
& $\tilde b_{\vec 0}^{*}\tilde c_{\vec 0}
+\tilde c_{\vec 0}^{*}\tilde b_{\vec 0}$
& ---\\

$\rho^{\mathrm{tp}}_2$
& $\eta^2$
& $-1/3$
& ---
& $\checkmark$ \\

$\rho^{\mathrm{tp}}_0$
& $\eta^0$
& $-1$
& $|\tilde b_{\vec 0}|^2$
& $\checkmark$ \\
\hline
\hline
\end{tabular}
\caption{Scaling, equation of state, and origin of the different topology-dependent contributions in Eq.~\eqref{Tmunu TP}.} 
\label{tab:topology_contributions} 
\end{table}
Focusing on an inflationary de Sitter expanding phase, in Fig. \ref{fig:plotLog} we choose $H=10^{14}\, \mathrm{GeV}$, corresponding to the GUT scale and $L=L_{\mathrm{LSS}}=4\cdot 10^{42}\,\mathrm{GeV}^{-1}$, corresponding to the diameter of the last scattering surface today. We normalize the scale factor to one today and, assuming temperatures $T_{\mathrm{t}}=10^{-13}\,\mathrm{GeV}$ today and $T_{\mathrm{reh}}=10^{15}\,\mathrm{GeV}$ at reheating, we find that inflation ends at $\eta_{\mathrm{end}}=-10^{14}\mathrm{GeV}^{-1}$. We then consider that inflation lasts $70$ e-folds, giving the conformal-time range
$-e^{70}10^{14}\,\mathrm{GeV}^{-1}<\eta<-10^{14}\,\mathrm{GeV}^{-1}$.
We take the overbarred contribution to the constant term, $\xi$, $\tilde{b}_{\vec 0}$, and $\tilde{c}_{\vec 0}$ to be of order unity  and we use the Wronskian condition in Eq. \eqref{Wronskian} to eliminate one of the four real parameters contained in $\tilde{b}_{\vec 0}$ and $\tilde{c}_{\vec 0}$. We do not aim here to explore the full parameter space; these choices are meant to illustrate how the different topology-dependent contributions evolve relative to one another during the de Sitter phase. Finally, we numerically evaluate the sum in Eq. \eqref{brakets} obtaining $B_1\simeq -3$. \\

Fig. \ref{fig:plotLog} shows that the total energy density is dominated over almost the entire plotted interval by the constant contribution, $\rho_{0}^{\mathrm{tp}}$, while, at the earliest times, the leading time-dependent contribution is $\rho_{6}^{\mathrm{tp}}(\eta)$, the $\eta^{6}$ zero-mode term. Neglecting the intermediate powers of $\eta$, the crossover to the constant-dominated regime may be estimated by
$|\rho_{6}^{\mathrm{tp}}(\eta_{\mathrm{cross}})|
=|\rho_{0}^{\mathrm{tp}}|$, which gives
\begin{equation}\label{eq:eta-cross}
    |\eta_{\mathrm{cross}}|  =   L   \left[  \frac{18|\rho_{0}^{\mathrm{tp}}|}   {|9-42\xi|H^{4}|\tilde c_{\vec 0}|^{2}}   \right]^{1/6} \,.
\end{equation}
For $|\eta|>|\eta_{\mathrm{cross}}|$, the $\eta^{6}$ contribution is larger, whereas for $|\eta|<|\eta_{\mathrm{cross}}|$ the constant sector dominates. For fixed $|\tilde c_{\vec 0}|$ and $|\xi|$, increasing $|\tilde b_{\vec 0}|$ shifts the constant-dominated regime toward earlier times because of the $\tilde b_{\vec 0}$ dependence in Eq. \eqref{Tmunu TP}.
At earlier times, the $\eta^{6}$ sector dominates relative to the other topology-dependent contributions and, since this term has equation of state $P_{6}^{\rm tp}=\rho_{6}^{\rm tp}$, its inclusion in the semiclassical Einstein equations generally drives the geometry away from the exact de Sitter solution. The curvature coupling in Eq. \eqref{action} then gives the homogeneous mode the time-dependent effective mass
\begin{equation}\label{eq:meff-definition}
    m_{\rm eff}^{2}(\eta)  =  \xi\left[R(\eta)-12H^{2}\right]\,.
\end{equation}
Quantifying the magnitude and sign of Eq. \eqref{eq:meff-definition}
when the backreaction of $\rho_{6}^{\rm tp}$ produces a substantial departure from de Sitter, would require solving the semiclassical evolution together with the field modes and recomputing the vacuum stress-energy tensor on the corrected geometry. We therefore do not proceed here to determine the complete time dependence or the sign of $m_{\rm eff}^{2}$ in the early regime. Nevertheless, the early time-dependent regime can provide the nonzero effective mass anticipated in Sec. \ref{Is the zero mode a particle?}. When the effective mass is positive, the homogeneous mode can admit a particle interpretation at early times and subsequently evolve into a massless zero-frequency degree of freedom. This provides a qualitative realization of the mechanism discussed in Ref. \cite{Ford:1989mf}.\\

We next consider the regime in which the explicitly time-dependent contributions in Eq. \eqref{Tmunu TP} are subleading, so that the energy density is effectively controlled by $\rho_{0}^{\rm tp}$, the $\eta$-independent part. In this constant-dominated regime the first Friedmann equation simplifies to 
\begin{equation}\label{eq:friedmann1}
    3 H^2 - \Lambda_{\rm eff} = - 24 \pi G  H^4 \xi \left[  |\tilde{b}_{\vec 0}|^2 + \frac{1}{16 \pi^3 }  \overbracket[0.8pt][0pt]{\sum_{\vec{n} \neq 0}} \frac{1}{|\vec n|^3}   \right] 
\end{equation}
and the equation of state of the topology-dependent vacuum stress tensor is
\begin{equation}
    P^{\rm tp}= - \rho^{\rm tp} \, .
\end{equation}
We first solve Eq. \eqref{eq:friedmann1} perturbatively in $G$, assuming a nonzero cosmological constant. We consider the expansion
\begin{equation}
    H = H_0 + G H_1 + \mathcal{O}(G^2) \,,
\end{equation}
and, substituting this into Eq. \eqref{eq:friedmann1} and keeping terms up to first order in $G$, we obtain
\begin{equation}\label{eq:semiclassical_friedmann}
    3 H_0^2 + 6 G H_0 H_1 - \Lambda_{\rm eff} = -24 \pi G H_0^4 \xi\left[ |\tilde{b}_{\vec 0}|^2+ \frac{ 1}{16 \pi^3 } \overbracket[0.8pt][0pt]{\sum_{\vec{n} \neq 0}} \frac{1}{|\vec n|^3}   \right] \, .
\end{equation}
Solving order by order in $G$ gives
\begin{equation}
    H_0 = \sqrt{\frac{\Lambda_{\rm eff}}{3}} \, , \qquad  H_1 =- 4 \pi \xi H_0^3 \left[ |\tilde{b}_{\vec 0}|^2 + \frac{ 1}{16 \pi^3} \overbracket[0.8pt][0pt]{\sum_{\vec{n} \neq 0}} \frac{1}{|\vec n|^3}   \right] \, .
\end{equation}
Therefore, for $H_0 \neq 0$, we find
\begin{equation}\label{eq:Hobs_nonzero_Lambda}
    H = \sqrt{\frac{\Lambda_{\rm eff}}{3}} - 4 \pi G \xi \left( \frac{\Lambda_{\rm eff}}{3}\right)^\frac{3}{2}\left[|\tilde{b}_{\vec 0}|^2 + \frac{  1 }{16 \pi^3}  \overbracket[0.8pt][0pt]{\sum_{\vec{n} \neq 0}} \frac{1}{|\vec n|^3} \right] + \mathcal{O}(G^2)\, .
\end{equation}
In this regime, the zero-mode contribution does not replace the cosmological constant sourcing a de Sitter inflationary phase, but rather shifts the Hubble rate. Whether the topology-dependent constant sector increases or decreases the expansion rate depends on the sign of $\xi$ and on the relative magnitude and sign of the overbarred contribution. The sign of the overbarred contribution cannot be determined in general since it depends on the shape of the fundamental domain.\\

We now consider the case in which the topology-dependent contribution is sourcing the de Sitter expansion alone, thereby considering $\Lambda_{\rm eff} = 0$ in Eq. \eqref{action}. From Eq. \eqref{eq:friedmann1} we obtain the solution
\begin{equation}\label{eq:Hobs_zero_mode}
    H^2  = -\frac{1}{8 \pi G \xi}\left[ \frac{  1}{16 \pi^3} \overbracket[0.8pt][0pt]{\sum_{\vec{n} \neq 0}} \frac{1}{|\vec n|^3}+    |\tilde{b}_{\vec 0}|^2 \right]^{-1} \, .
\end{equation}
We note that for $\xi<0$, an expanding solution can be supported provided the total contribution in parentheses is positive. For $\xi>0$, an expanding solution is possible only if the overbarred contribution is negative and sufficiently large in magnitude to overcome the positive zero-mode term. If both $\xi$ and the overbarred contribution to the constant term are positive, the zero-mode term cannot source an expanding de Sitter solution.  \\

We estimate the values of the zero-mode integration constant $|\tilde{b}_{\vec 0}|$ required for the topology-dependent stress-energy tensor to source  an inflationary de Sitter phase with $H =H_{\rm inf}$ without a separate cosmological-constant term in the action. We find that this requirement can be reformulated as a condition on the choice of vacuum state. By setting $H =H_{\rm inf}=10^{14} $ GeV in Eq. \eqref{eq:Hobs_zero_mode} we find
\begin{equation}\label{b02}
    \left[ - \xi |\tilde{b}_{\vec 0}|^2 - \frac{ \xi}{16 \pi^3} \overbracket[0.8pt][0pt]{\sum_{\vec{n} \neq 0}} \frac{1}{|\vec n|^3} \right] \simeq 6 \cdot 10^{8}\, . 
\end{equation}
If $\xi>0 $, the zero-mode term decreases the quantity in square brackets, so these large positive values cannot be obtained from topology unless the overbarred contribution is negative and extremely large in magnitude. For $\xi <0 $ the required value can instead be obtained through a sufficiently large zero-mode amplitude. Assuming a single dominant species with $| \xi|  \sim 1$ and the overbarred contribution to the constant term of order unity this corresponds to $| \tilde{b}_{\vec 0}| \sim 10^{4} $. For several independent species, the quantity in square brackets generalizes schematically to a sum over species,
\begin{equation}
   \left[  -   \xi |\tilde{b}_{\vec 0}|^2- \frac{ \xi}{16 \pi^3} \overbracket[0.8pt][0pt]{\sum_{\vec{n} \neq 0}} \frac{1}{|\vec n|^3} \right] \quad \longrightarrow \quad \sum_s\left[  -   \xi_s |\tilde{b}_{\vec 0}^{(s)}|^2 - \frac{ \xi_s}{16 \pi^3} \overbracket[0.8pt][0pt]{\sum_{\vec{n} \neq 0}} \frac{1}{|\vec n|^3} \right]  \, .
\end{equation}
A large effective zero-mode amplitude could therefore arise collectively. \\

Our results show that, in a toroidal universe, the topology-dependent part of the vacuum stress-energy tensor of a massless scalar field can have nontrivial effects on an early-Universe de Sitter expanding phase. When a nonzero $\Lambda_{\rm eff}$ already sources the de Sitter background, the topology-dependent contribution shifts the corresponding Hubble rate. Conversely, when $\Lambda_{\rm eff}=0$, the topology-dependent vacuum stress-energy tensor can itself support an expanding de Sitter solution. In the regime relevant for most of the inflationary evolution considered here, the stress-energy tensor is dominated by its constant contribution, including the zero-mode term proportional to $|\tilde b_{\vec 0}|^2$, which has equation of state $P=-\rho$. Requiring this contribution to source a GUT inflationary expansion with $H=H_{\rm inf}\sim10^{14}\,{\rm GeV}$ translates into a condition on the IR structure of the vacuum state; for a single massless field with $|\xi|\sim1$ and a negligible nonzero-mode contribution to the constant sector, we find $|\tilde b_{\vec 0}|\sim10^4$. Since the dependence of the integration constants on the underlying parameters, such as the scale of topology, is not a priori known, our estimates do not by themselves determine the degree of tuning involved.\\

We recall that the apparent absence of an explicit $L$-dependence in the zero-mode term results from the rescaling $ b_{\vec 0}=(HL)^{3/2}\tilde b_{\vec 0}$ introduced in Eq. \eqref{rescaling}. Before this rescaling, the zero-mode contribution contains the explicit volume factor $|b_{\vec 0}|^2/L^3$, while the integration constant $b_{\vec 0}$ may itself also depend on $L$ through the choice of quantum state. Thus, the rescaling does not remove the explicit dependence on the topology scale, but rather absorbs its into $\tilde b_{\vec 0}$. By contrast, the constant nonzero-mode contribution is independent of an overall rescaling of the fundamental domain and depends only on its dimensionless shape parameters. Our estimates nevertheless show that the relevant cosmological scales are encoded in the IR structure of the state. We do not attempt here to identify a preferred region of parameter space or to construct a complete inflationary model realizing this mechanism. Rather, our aim is to demonstrate how the topology-induced vacuum contribution can enter the semiclassical dynamics as well as illustrate the range of effects allowed by the remaining IR freedom of the state. We find that, in a spacetime with compact spatial sections, the effective source driving de Sitter expansion depends not only on local terms in the action, but also on the choice of quantum vacuum state compatible with the global topology. \\

This analysis can be extended beyond the inflationary de Sitter phase by propagating the zero-mode contribution through the subsequent cosmological evolution. This requires solving the homogeneous mode equation separately during radiation and matter domination and during a later accelerating phase, while matching the mode function and its first derivative across the transitions. Throughout the cosmological expansion history, the zero-mode contribution to the stress-energy tensor does not retain the same scaling behavior found during the inflationary de Sitter phase, since its evolution depends on the background expansion and on the mode coefficients inherited from the preceding epoch. During a late-time de Sitter phase, the behavior of the zero mode is controlled by the nonminimal coupling and the curvature scale entering the definition of the shifted physical mass $m_{\rm p}$, and can again contain a contribution that approaches a constant energy density with $P=-\rho$. The late-time zero-mode sector can therefore provide a cosmological-constant contribution, whose amplitude is inherited from its evolution through the preceding radiation- and matter-dominated epochs.

\section{Conclusion}\label{Conclusion}

In this work, we studied how nontrivial spatial topology affects the choice of quantum vacuum in QFT on curved spacetime. Using locally de Sitter spacetime with toroidal spatial sections as a representative example, we first confirmed that the Hadamard conditions remain insensitive to global topology. This is expected, since nontrivial topology leaves the local UV structure unchanged. By contrast, the IR sector of the theory is affected by the identifications defining the underlying topology and breaking global de Sitter invariance. Consequently, the usual symmetry argument used in the simply connected case to discard the zero mode no longer applies. We treated the zero mode as a genuine IR freedom of the vacuum state: since the Hadamard conditions leave a set of physically admissible vacua, we did not attempt to introduce a new prescription that uniquely fixes such remaining freedom. Instead, we took a phenomenological approach to studying the consequences of the topology-induced zero-mode freedom in semiclassical gravity. We highlight that the zero-mode contribution studied in this work is qualitatively different from particle production. It is not associated with excitations above a chosen vacuum, nor with a particle-like fluid created by the time dependence of the background \cite{Kolb:2023ydq,Ford:2021syk}. It is instead a Casimir-type vacuum contribution \cite{Casimir:1948dh,DeWitt:1979dd}, and nontrivial topology, by breaking global de Sitter invariance, removes the usual symmetry criterion that would otherwise be used to eliminate this contribution. This furthermore has a broader cosmological relevance: realistic inflationary backgrounds are only approximately de Sitter, so exact de Sitter invariance cannot in general be invoked to uniquely select the vacuum. Our results therefore raise the question of how a zero-mode-free, BD-like vacuum is selected when the Universe begins in an approximately de Sitter regime, and what physical principle determines this choice.\\

We found that the zero-mode sector can contain a contribution with equation of state $P=-\rho$, contributing to the semiclassical Friedmann equation as an effective cosmological constant source. Our estimates show that a de Sitter inflationary expansion can be sourced by the topology-dependent contribution alone, for a sufficiently large zero-mode amplitude. We thereby demonstrate that, in a universe with nontrivial spatial topology, the effective source driving a de Sitter expansion is not determined solely by parameters in the action, but also depends on topology-dependent IR contributions to the quantum vacuum state. In this sense, the vacuum-selection ambiguity that remains once the usual de Sitter-invariance argument is no longer available to discard the zero-mode contribution becomes directly relevant to the semiclassical cosmological dynamics.\\

Although we restricted our analysis to an early de Sitter phase, the zero-mode need not cease to be relevant once inflation ends. Its contribution can be evolved through the subsequent radiation- and matter-dominated epochs and into a later de Sitter expanding phase. The corresponding stress-energy contribution evolves nontrivially across these epochs and the equation of state during the late de Sitter phase depends on the scalar mass, the nonminimal coupling, and the curvature scale entering the definition of the shifted physical mass $m_{\rm p}$. For appropriate choices of these parameters, one of the late-time zero-mode contributions can again approach a constant energy density with equation of state $P=-\rho$. The IR freedom identified here may therefore have implications beyond the early-Universe de Sitter phase, including contributing to the cosmological constant at late times. However, establishing whether the zero-mode contributions presented in this work can naturally evolve into a contribution to the cosmological constant at late times requires following their dynamics and backreaction throughout the full cosmological history, which we will explore in future work.\\

Finally, while the torus provides a setting in which the argument can be made explicit, the presented mechanism applies to any of the nontrivial Euclidean topologies ($E_2 \, \cdots E_{17}$), since they are also generated by global identifications that reduce the symmetry of the simply connected covering space. These topologies have a rich set of shape parameters \cite{COMPACT:2022nsu,COMPACT:2023rkp,COMPACT:2025adc}, and their compactification scales can span many orders of magnitude, with some directions potentially approaching microscopic scales \cite{InPreparation}. Moreover, different topologies will in general lead to different Casimir contributions. The resulting topology-dependent vacuum stress tensor can therefore depend sensitively on both the topological scales and the parameters characterizing the identifications. This enlarged parameter space could significantly modify both the magnitude and structure of the topology-induced vacuum contribution. Determining how these possibilities affect early-Universe de Sitter dynamics, and whether they can generate phenomenologically relevant vacuum contributions at later times, is left for future work. \\

\section*{Acknowledgments}
We thank L. Knox, B. Freivogel and B. Fornal for helpful discussions during the development of this work.
A.N.\ is supported by the Richard S.\ Morrison Fellowship.  
Y.A.\ acknowledges support by the Spanish Research Agency (Agencia Estatal de Investigaci\'on)'s grant RYC2020-030193-I/AEI/10.13039/501100011033, by the European Social Fund (Fondo Social Europeo) through the Ram\'{o}n y Cajal programme within the State Plan for Scientific and Technical Research and Innovation (Plan Estatal de Investigaci\'on Cient\'ifica y T\'ecnica y de Innovaci\'on) 2017-2020, by the Spanish Research Agency through the grant IFT Centro de Excelencia Severo Ochoa No CEX2020-001007-S funded by MCIN/AEI/10.13039/501100011033, by the Spanish National Research Council (CSIC) through the Talent Attraction grant 20225AT025, and by the Spanish Research Agency's Consolidaci\'on Investigadora 2024 grant CNS2024-154430.
J.C.D.\ is supported by the Spanish Research Agency (Agencia Estatal de Investigaci\'on), the Ministerio de Ciencia, Innovaci\'on y Universidades, and the European Social Funds through grant JDC2023-052152-I, as part of the Juan de la Cierva programme. 
M.M.B. acknowledges support by the Spanish Ministry of Science, Innovation and Universities under the FPU predoctoral grant FPU22/02306. 
A.T.\ is supported by the European Union's Horizon Europe research and innovation 
programme under the Marie Sk\l{}odowska-Curie grant agreement No.\ 101126636. 

\clearpage 

\appendix

\section{Improved stress-energy tensor}
\label{DetailsImprov}

We begin by recalling the vacuum stress-energy tensor associated with the standard action for a nonminimally coupled scalar field,
\begin{equation}
    S  = \frac{1}{2}  \int d^4x \sqrt{-g}  \left[   -\nabla_\sigma\phi\nabla^\sigma\phi  -m^2\phi^2  -\xi R\phi^2  \right] \, .
\end{equation}
The corresponding equation of motion is
\begin{equation}\label{EOMStandard}
    \left(\nabla_\sigma \nabla^\sigma -m^2-\xi R\right)\phi=0 
\end{equation}
and, on a locally de Sitter background, the stress energy tensor can be written as
\begin{equation}\label{Tmunugeneral1}
    \begin{aligned}
        \langle T_{\alpha\beta}\rangle   =  \left\langle
        \nabla_\alpha\phi\nabla_\beta\phi   -\frac{1}{2}g_{\alpha\beta} \nabla_\sigma \phi \nabla^\sigma \phi -\frac{1}{2}m^2g_{\alpha\beta}\phi^2 -\xi  \left(  \nabla_\alpha\nabla_\beta  -g_{\alpha\beta}\nabla_\sigma \nabla^\sigma +3H^2g_{\alpha\beta}  \right)\phi^2 \right\rangle  \, .
    \end{aligned}
\end{equation}
In this parametrization, the same coupling $\xi$ enters both the field equation in Eq. \eqref{EOMStandard}, and hence the mode functions, and the stress-energy tensor in Eq. \eqref{Tmunugeneral1}. \\

In the main text, however, we wish to separate these two roles of $\xi$. We therefore introduce the action
\begin{equation}\label{actionshifted}
    S  =  \frac{1}{2} \int d^4x \sqrt{-g}
    \left[  -\nabla_\sigma\phi\nabla^\sigma \phi  -m_{\rm p}^2\phi^2  -\xi\left(R-12H^2\right)\phi^2   \right] \, ,
\end{equation}
here $m_{\rm p}$ denotes the physical mass appearing in the mode equation on the reference de Sitter background. Since $R=12H^2$ on this background, the field equation reduces to
\begin{equation}
    \left(\nabla_\sigma \nabla^\sigma-m_{\rm p}^2\right)\phi=0 
\end{equation}
and the parameter $\xi$ no longer shifts the mode equation. Therefore, such shifted form is useful because on the reference de Sitter background the curvature-dependent term drops out of the field equation and $m_{\rm p}$ is directly the mass entering the mode functions. Varying the action with respect to the metric gives
\begin{equation}\label{Tappendix}
    \begin{aligned}
        \langle T_{\alpha\beta}\rangle  = \left\langle
        \left(1-2\xi\right)\nabla_\alpha\phi\nabla_\beta\phi  -\left(\frac{1}{2}-2\xi\right) g_{\alpha\beta}\nabla_\sigma\phi\nabla^\sigma\phi   -2\xi\phi\nabla_\alpha\nabla_\beta\phi  +2\xi g_{\alpha\beta} \phi \nabla_\sigma \nabla^\sigma \phi  \right\rangle
        \\
        + \left\langle -\frac{1}{2}   \left(m_{\rm p}^2-12\xi H^2\right) g_{\alpha\beta}\phi^2  +
        \xi \left( R_{\alpha\beta} -\frac{1}{2} g_{\alpha\beta}R \right)\phi^2 \right\rangle 
    \end{aligned}
\end{equation}
which, in de Sitter, results 
\begin{equation} \label{TmunuImpr}
    \begin{aligned}
        \langle T_{\alpha\beta}\rangle    =  \left\langle \nabla_\alpha\phi\nabla_\beta\phi  -\frac{1}{2}g_{\alpha\beta} \nabla_\sigma \phi\nabla^\sigma \phi   -\frac{1}{2}m_{\rm p}^2g_{\alpha\beta} \phi^2 -\xi
        \left( \nabla_\alpha \nabla_\beta  -g_{\alpha\beta} \nabla_\sigma \nabla^\sigma -3H^2g_{\alpha\beta}
        \right)\phi^2 \right\rangle \, .
    \end{aligned}
\end{equation}
Relative to Eq. \eqref{Tmunugeneral1}, the sign of the final curvature term is reversed. This is the form used throughout the main text.\\

It is useful to clarify the sense in which the $\xi$-dependent contribution in Eq. \eqref{TmunuImpr} constitutes an improvement term\footnote{In flat spacetime, one may add the identically conserved derivative term $\Delta T_{\alpha\beta} = (\partial_\alpha\partial_\beta -\eta_{\alpha\beta}\partial_\sigma \partial^\sigma)\phi^2$  to the scalar field stress-energy tensor. For fields that decay sufficiently rapidly at spatial infinity, or on compact spatial slices without boundary, this term does not change the total energy or momentum.}. Define
\begin{equation}\label{Improvement}
    \Delta T_{\alpha\beta}    =   \left(   \nabla_\alpha\nabla_\beta  -g_{\alpha\beta}\nabla_{\sigma}  \nabla^{\sigma}  -3H^2g_{\alpha\beta}\right)\phi^2 \, ,
\end{equation}
then
\begin{equation}
    T_{\alpha\beta} = T^{(0)}_{\alpha\beta}-\xi \Delta T_{\alpha\beta} \,,
\end{equation}
where $T^{(0)}_{\alpha\beta}$ is the minimally coupled stress-energy tensor with mass $m_{\rm p}$. $\Delta T_{\alpha\beta}$ is identically conserved on de Sitter. Indeed,
\begin{equation}
        \nabla^\alpha \Delta T_{\alpha\beta}  = \nabla^\alpha\nabla_\alpha\nabla_\beta \phi^2  -  \nabla_\beta \nabla_{\alpha} \nabla^{\alpha} \phi^2  - 3H^2\nabla_\beta\phi^2 =  R_{\beta\alpha}\nabla^\alpha\phi^2 - 3H^2\nabla_\beta\phi^2 =  0 \, ,
\end{equation}
where we used
\begin{equation}
    g^{\alpha\sigma} \left[  \nabla_\alpha,\nabla_\beta \right]\nabla_\sigma\phi^2  =R_{\sigma\beta}\nabla^\sigma\phi^2\, ,
\end{equation}
together with $R_{\alpha\beta}=3H^2g_{\alpha\beta}$. Thus, the term
$-3H^2g_{\alpha\beta}\phi^2$ is precisely the curvature correction required to make the improvement tensor identically conserved on de Sitter.\\

From Eq. \eqref{TmunuImpr}, the energy density is
\begin{equation}\label{rhoIMP}
    \rho  = - \langle T_0{}^0\rangle = \frac{1}{a^2} \left\langle  \frac{1}{2}\phi^{\prime\,2} +  \frac{1}{2}(\nabla_i\phi)^2  +\frac{a^2}{2}m_{\rm p}^2\phi^2- \xi \left(  \nabla_0\nabla_0  + a^2\nabla_\alpha \nabla^\alpha  + 3H^2a^2 \right)\phi^2 \right\rangle \, ,
\end{equation}
while the mixed spatial components of the stress tensor are
\begin{equation}\label{PIMP}
    \begin{aligned}
        P_i{}^j =\langle T_i{}^j\rangle= \left\langle   \nabla_i\phi\nabla^j\phi + \frac{1}{2a^2}\delta_i{}^j  \left[  \phi^{\prime\,2}   -  (\nabla_k\phi)^2 \right]  - \frac{1}{2}m_{\rm p}^2\phi^2\delta_i{}^j  -  \xi   \left( \nabla_i\nabla^j  -  \delta_i{}^j\nabla_\alpha \nabla^\alpha - 3H^2\delta_i{}^j\right)\phi^2  \right\rangle \, .
    \end{aligned}
\end{equation}
Equations \eqref{rhoIMP} and \eqref{PIMP} are equivalent to those obtained from Eq. \eqref{Tmunugeneral1} upon using
\begin{equation}\label{replacement}
    m^2=m_{\rm p}^2-12\xi H^2 \,.
\end{equation}
In this shifted parametrization, the curvature-dependent terms combine such that Eq. \eqref{Tmunugeneral1} takes the form of Eq. \eqref{TmunuImpr}. Conversely, the standard parametrization is recovered through $m_{\rm p}^2=m^2+12\xi H^2$. In particular, the standard massless conformally coupled case, $m=0$ and $\xi=1/6$, corresponds to $m_{\rm p}^2=2H^2$, whereas $m_{\rm p}=0$ describes a mode equation that is massless on the reference de Sitter background independently of $\xi$.

\subsection{Nonzero-mode contribution}
\label{NonZero}

We now specialize Eqs. \eqref{rhoIMP} and \eqref{PIMP} to the nonzero-mode sector of the field. In conformal coordinates, the energy density takes the form
\begin{equation}\label{rhoNZpre}
    \begin{aligned}
       \rho   = \frac{1}{a^2} \left\langle  \frac{1}{2}\phi^{\prime\,2}  +  \frac{1}{2}(\partial_{x_i}\phi)^2   +   \frac{a^2}{2}m_{\rm p}^2\phi^2  -  \xi
       \left(  -3\frac{a'}{a}\partial_\eta +  \partial_{x_i}^2  +  3H^2a^2 \right)\phi^2  \right\rangle \, .
    \end{aligned}
\end{equation}
Similarly, the mixed spatial components are
\begin{equation}\label{PNZpre}
    \begin{aligned}
        P_i{}^j  = \frac{1}{a^2}  \left\langle 
        \partial_{x_i}\phi\,\partial^{x_j}\phi  +
        \frac{1}{2}\delta_i{}^j  \left[  \phi^{\prime\,2}   -  (\partial_{x_k}\phi)^2   \right]  -  \frac{a^2}{2}m_{\rm p}^2 \phi^2\delta_i{}^j
        \right\rangle
        \\
        -  \frac{\xi}{a^2}  \left\langle   \left[  \partial_{x_i}\partial^{x_j}   + \left(  \partial_\eta^2  -  \partial_{x_k}^2  +   \frac{a'}{a}\partial_\eta  -  3H^2a^2  \right)\delta_i{}^j   \right]\phi^2   \right\rangle \, .
    \end{aligned}
\end{equation}
Here, $\partial^{x_j}\equiv\delta^{jk}\partial_{x_k}$, so that spatial indices on coordinate derivatives are raised with $\delta^{ij}$. To evaluate the derivative terms in the improvement contribution, we use
\begin{equation} \label{Ids1}
    \begin{aligned}
        \partial_\eta \phi^2 = \phi\phi' + \phi'\phi , \quad \partial_\eta^2\phi^2  =\phi\phi''+\phi''\phi + 2\phi^{\prime\,2},  \quad \partial_{x_i}\partial^{x_j}\phi^2=\phi\,\partial_{x_i}\partial^{x_j}\phi+\partial_{x_i}\partial^{x_j}\phi\,\phi+2\partial_{x_i}\phi\,\partial^{x_j}\phi \, ,
    \end{aligned}
\end{equation}
together with the nonzero-mode expansion
\begin{equation}\label{modeExpansionNZ}
    \phi(\eta,\vec{x})  =  \frac{1}{\sqrt{\mathrm{Vol}}}  \sum_{\vec{n}\neq\vec{0}}   \left[  a_{\vec{n}}\,  e^{i\vec{k}_{\vec{n}}\cdot\vec{x}}  \varphi_{\vec{n}}(\eta)  + a_{\vec{n}}^\dagger\,  e^{-i\vec{k}_{\vec{n}} \cdot\vec{x}} \varphi_{\vec{n}}^*(\eta)  \right]\,,    
\end{equation}
where
\begin{equation}\label{modeFunctionNZ}
    \varphi_{\vec{n}}(\eta)  =  \frac{H\sqrt{\pi|\eta|^3}}{2}
    e^{-i\frac{\pi}{2}\left(\nu+\frac{1}{2}\right)} H_{\nu}^{(1)} \left(  -|\vec{k}_{\vec{n}}|\eta
    \right) \,,  \qquad   \vec{n}\neq\vec{0} \,.
\end{equation}
Substituting Eqs. \eqref{Ids1}-\eqref{modeFunctionNZ} into Eqs. \eqref{rhoNZpre} and \eqref{PNZpre}, and using standard Hankel function recurrence relations, we obtain
\begin{equation}
    \begin{aligned}\label{rhoMod}
       \rho &= \frac{ \pi H^4 |\eta|^3}{8 \mathrm{Vol}} \left[ \left(\frac{3}{2}+\nu \right)^2+\frac{m_{\rm p}^2}{H^2} \right] \sum_{\vec{n} \neq 0} |H_{\nu}^{(1)}(-|\vec{k}_n| \eta )|^2 \\
       &+ \frac{ \pi H^4 |\eta|^5}{8 \mathrm{Vol}} \sum_{\vec{n} \neq 0} |\vec{k}_n|^2 \left[ |H_{\nu+1}^{(1)}(-|\vec{k}_n| \eta )|^2+|H_{\nu}^{(1)}(-|\vec{k}_n| \eta )|^2  \right] \\
        & -\frac{\pi H^4|\eta|^4}{8\mathrm{Vol}}  \left(\frac{3}{2}+\nu \right) \sum_{\vec{n} \neq 0}|\vec{k}_n| \left[H_{\nu}^{(1)}(-|\vec{k}_n| \eta )  H_{\nu+1}^{(2)}(-|\vec{k}_n| \eta )+ H_{\nu}^{(2)}(-|\vec{k}_n| \eta )  H_{\nu+1}^{(1)}(-|\vec{k}_n| \eta )\right]\\ 
       & -\frac{\pi \xi H^4|\eta|^3}{8 \mathrm{Vol}} \sum_{\vec{n} \neq 0}\left[6 |\vec{k}_n| \eta \left(H_{\nu}^{(1)}(-|\vec{k}_n| \eta) H_{\nu+1}^{(2)}(-|\vec{k}_n| \eta)+H_{\nu}^{(2)}(-|\vec{k}_n| \eta) H_{\nu+1}^{(1)}(-|\vec{k}_n| \eta)\right)\right. \\
       &\left.+12 (\nu+2)\left|H_{\nu}^{(1)}(-|\vec{k}_n| \eta)\right|^2\right] \, ,
    \end{aligned}
\end{equation}
\begin{equation}\label{PMod}
    \begin{aligned}
        P_{i}{}^{j}=& \frac{\pi H^4|\eta|^5}{4 \mathrm{Vol}} \sum_{\vec{n} \neq 0} (\vec{k}_n)_i (\vec{k}_n)^{j} |H_{\nu}^{(1)}(-|\vec{k}_n| \eta )|^2 + \delta_{i}{}^{j} \left\{\frac{ \pi H^4 |\eta|^3}{8 \mathrm{Vol}} \left[ \left(\frac{3}{2}+\nu \right)^2 -\frac{m_{\rm p}^2}{H^2} \right]\sum_{\vec{n} \neq 0} |H_{\nu}^{(1)}(-|\vec{k}_n| \eta )|^2 \right. \\
        &  + \frac{ \pi H^4 |\eta|^5}{8 \mathrm{Vol}} \sum_{\vec{n} \neq 0} |\vec{k}_n|^2 \left[ |H_{\nu+1}^{(1)}(-|\vec{k}_n| \eta )|^2 -|H_{\nu}^{(1)}(-|\vec{k}_n| \eta )|^2 \right] \\
        & \left .-\frac{\pi H^4|\eta|^4}{8\mathrm{Vol}}  \left(\frac{3}{2}+\nu \right) \sum_{\vec{n} \neq 0}|\vec{k}_n| \left[H_{\nu}^{(1)}(-|\vec{k}_n| \eta )  H_{\nu+1}^{(2)}(-|\vec{k}_n| \eta )+ H_{\nu}^{(2)}(-|\vec{k}_n| \eta )  H_{\nu+1}^{(1)}(-|\vec{k}_n| \eta )\right] \right\}\\
        &- \frac{\pi \xi H^4|\eta|^3}{4 \mathrm{Vol}} \sum_{\vec{n} \neq 0}  {\left[\left(4 \nu(\nu+2)-2|\vec{k}_n|^2\eta^2\right)\left|H_{\nu}^{(1)}(-|\vec{k}_n| \eta)\right|^2\right.} \\ 
        &+2(\nu+2)|\vec{k}_n|\eta\left(H_{\nu}^{(1)}(-|\vec{k}_n| \eta) H_{\nu+1}^{(2)}(-|\vec{k}_n| \eta)+H_{\nu}^{(2)}(-|\vec{k}_n| \eta) H_{\nu+1}^{(1)}(-|\vec{k}_n| \eta)\right) \\ 
        &\left.+2|\vec{k}_n|^2 \eta^2\left|H_{\nu+1}^{(1)}(-|\vec{k}_n| \eta)\right|^2\right]\delta_i{ }^j \,.
    \end{aligned}
\end{equation}
In the physical massless limit, $m_{\rm p}^2=0$, one has $\nu = \frac{3}{2}$; using the explicit expressions for $H^{(1)}_{\frac{3}{2}}$, $H^{(1)}_{\frac{5}{2}}$, Eqs. \eqref{rhoMod} and \eqref{PMod} reduce to Eq. \eqref{eq:Components1} of the main text.\\

Equations \eqref{rhoMod} and \eqref{PMod} are written in the shifted parametrization introduced in Eq. \eqref{actionshifted}, but they are equivalent to the expressions we would obtain from Eq. \eqref{Tmunugeneral1} after making the replacement in Eq. \eqref{replacement}. The expressions in the standard parametrization are recovered by setting $m_{\rm p}^2=m^2 + 12 \xi H^2$. 

\clearpage

\section{Extracting contributions of nontrivial topology}\label{PoissonEwald}

In this appendix, we detail the steps leading from Eq. \eqref{eq:Components1} to the UV-finite, topology-dependent result in Eq. \eqref{Tmunu TP}.\\

Specializing Eq. \eqref{eq:Components1} to the cubic torus, for which $\vec{k}= \frac{2 \pi}{L} \vec{n}$, we obtain
\begin{equation}\label{eq:ComponentsApp}
    \begin{aligned}
        \rho^{\mathrm{imp}}_{\nu=\frac{3}{2}} = & \frac{\pi \eta^4 H^4}{ L^4}\sum_{\vec{n} \neq 0}  |\vec{n}|  + \left( 1 - 18 \xi \right) \frac{\eta^2 H^4}{8 \pi  L^2}  \sum_{\vec{n} \neq 0} \frac{1}{|\vec{n}|} - \frac{3 H^4\xi}{16 \pi^3 }\sum_{\vec{n} \neq 0} \frac{1}{|\vec{n}|^3} -\frac{3H\xi}{L^3}|b_{\vec 0}|^2  \\
        &+(9-42\xi)\frac{H^7|\eta|^6}{18 L^3}|c_{\vec 0}|^2  +\frac{4H^4\xi\eta^3}{L^3} \left(   b_{\vec 0}^*c_{\vec 0}+b_{\vec 0}c_{\vec 0}^* \right) \, ,\\
        P^{\mathrm{imp}}_{\nu=\frac{3}{2}}{}_i{}^j  = & \frac{\pi \eta^4 H^4}{3 L^4} \sum_{\vec n\neq 0}|\vec n|\delta_i{}^j + \left(18\xi-1\right) \frac{\eta^2 H^4}{24\pi L^2} \sum_{\vec n\neq 0}\frac{1}{|\vec n|} \delta_i{}^j+ \frac{3H^4\xi}{2(2\pi)^3} \sum_{\vec n\neq 0}\frac{1}{|\vec n|^3} \delta_i{}^j + \frac{3H\xi}{L^3}|b_{\vec 0}|^2 \delta_i{}^j \\
        & + (9-42\xi) \frac{H^7|\eta|^6}{18L^3} |c_{\vec 0}|^2 \delta_i{}^j \,.
    \end{aligned}
\end{equation}
The topology-dependent part of the stress-energy tensor can be extracted once the three sums
\begin{equation}\label{UVsums}
        \sum_{\vec n\neq0}|\vec n|\, ,\qquad \sum_{\vec n\neq0}\frac{1}{|\vec n|}\, ,\qquad \sum_{\vec n\neq0}\frac{1}{|\vec n|^3} 
\end{equation}
have been decomposed into their covering-space and finite-volume contributions. \\

For the first two sums in Eq. \eqref{UVsums} the decomposition is implemented by the Poisson summation formula, which rewrites a sum over the discrete torus momenta as a continuum integral, corresponding to the covering-space contribution, together with a sum over nonzero winding images, corresponding to the finite-volume, topology-dependent contribution. To apply the Poisson summation formula in Eq. \eqref{eq:Poisson}, we need to temporarily add the zero mode in the sum. For the sum proportional to $|\vec n|$, this is harmless because its zero-mode contribution vanishes. We therefore obtain
\begin{equation}\label{eq:sums1}
    \begin{aligned}
         \frac{2 \pi}{L^4} \sum_{\vec n\in\mathbb{Z}^3} |\vec n |
        &= \int \frac{d^3 q}{(2\pi)^3}|\vec q| -\frac{1}{\pi^2 L^4}  \sum_{\vec r\neq \vec 0} \frac{1}{|\vec r|^4} \, .
    \end{aligned}
\end{equation}
However, in general, by temporarily including the zero mode in the sum one introduces an artificial IR divergence. The Ewald summation method provides a prescription for isolating the finite contribution by rewriting the sum as follows (see Sec. \ref{Ewald} for details)
\begin{equation}\label{eq:sums2}
    \begin{aligned}
         \sum_{\vec{n} \neq 0} \frac{1}{|\vec{n}|}=  &2 \pi L^2 \int \frac{d^3 q}{(2 \pi)^3} \frac{1}{|\vec{q}|}+\frac{L^2}{\pi} \: \: \widetilde{\sum_{\vec{r} \neq 0}} \frac{1}{|\vec{l}_r|^2}\\
        =  &2 \pi L^2 \int \frac{d^3 q}{(2\pi)^3}\frac{1}{|\vec q|}+ \frac{1}{\pi} \sum_{\vec{r} \neq 0} \frac{e^{-\lambda L^2 |\vec{r}|^2}}{|\vec{r}|^2}+ \sum_{\vec{n} \neq 0} \frac{1}{|\vec{n}|} \operatorname{erfc}\left(\frac{\pi \left|\vec{n}\right|}{L \sqrt{\lambda}}\right)-\frac{2 \sqrt{\pi}}{L } \frac{1}{\sqrt{\lambda}}  - \frac{\lambda L^2}{\pi} \, ,
    \end{aligned}
\end{equation}
where the tilde over the image sum in the first line denotes the IR-finite part. The IR-finite image terms appearing in Eq. \eqref{eq:sums2} is extracted from the following Ewald identity
\begin{equation}\label{eq:ewaldSUms}
    \begin{aligned}
        &\sum_{\vec{r} \neq 0} \frac{1}{|\vec{l}_r|^2}   = \sum_{\vec{r} \neq 0} \frac{e^{-\lambda |\vec{l}_r|^2}}{|\vec{l}_r|^2}+\frac{2 \pi^2}{\operatorname{Vol}} \sum_{\vec{n} \neq 0} \frac{1}{|k_n|} \operatorname{erfc}\left(\frac{\left|k_n\right|}{2 \sqrt{\lambda}}\right)-\frac{2 \pi^{3 / 2}}{\operatorname{Vol} } \frac{1}{\sqrt{\lambda}} - \lambda  +\frac{2 \pi^{3 / 2}}{L^2 }\lim_{\epsilon \to 0} \frac{1}{\sqrt{\epsilon}}
    \end{aligned}
\end{equation}
where the parameter $\lambda>0$ appearing in Eqs. \eqref{eq:sums2} and \eqref{eq:ewaldSUms} is the Ewald splitting parameter. It has dimensions of inverse length squared, so that the natural dimensionless parameter is $\mu = \lambda L^2$. The complete expression does not depend on $\mu$; its sole role is to redistribute the same finite result among the real-space sum, the reciprocal-space sum, and the explicit subtraction terms.  Equivalently, differentiating the complete expressions with respect to $\mu$ gives zero after using the same Poisson, or theta-function, identity employed in the derivation (see Appendix \ref{Ewald} for more details). In practice, $\lambda$ may therefore be chosen to optimize the convergence of the sum. Subtracting the UV-divergent covering-space contributions appearing in Eqs. \eqref{eq:sums1} and \eqref{eq:sums2}, we obtain the topology-dependent parts of the sums:
\begin{equation}\label{eq:Sumsfinal}
    \begin{aligned}
    &\left[ \sum_{\vec n\in\mathbb{Z}^3} |\vec n | \right]^{\mathrm{tp}}  = -\frac{1}{2\pi^3 }
    \sum_{\vec r\neq \vec 0}
    \frac{1}{|\vec r|^4} \, ,\\
        &\left[\sum_{\vec{n} \neq 0} \frac{1}{|\vec{n}|} \right]^{\mathrm{tp}}  = \frac{1}{\pi} \sum_{\vec{r} \neq 0} \frac{e^{-\lambda L^2 |\vec{r}|^2}}{|\vec{r}|^2}+ \sum_{\vec{n} \neq 0} \frac{\operatorname{erfc}\left(\frac{\pi \left|\vec{n}\right|}{L \sqrt{\lambda}}\right)}{|\vec{n}|} -\frac{2 \sqrt{\pi}}{L } \frac{1}{\sqrt{\lambda}}  - \frac{\lambda L^2}{\pi} 
        \,.
    \end{aligned}
\end{equation}
The last expression in Eq. \eqref{eq:Sumsfinal} is precisely the quantity $B_1$ defined in Eq.~\eqref{brakets}. \\

For the third sum in Eq. \eqref{UVsums}, however, the same Ewald procedure cannot be used to extract the finite topology-dependent contribution \cite{NIJBOER1957309}. The artificial IR divergence introduced in applying the Poisson/Ewald decomposition cannot in this case be handled so as to unambiguously isolate the finite topology-dependent contribution. We therefore do not use the Ewald representation for this term. Instead, we rely on the general separation established in Sec. \ref{topology dep Tmunu} between the local, UV-divergent covering-space contribution and the finite, scheme-independent contribution associated with the nontrivial topology, and formally write
\begin{equation}
        \sum_{\vec n\neq0}\frac{1}{|\vec n|^3} = \left[ \sum_{\vec n\neq0}\frac{1}{|\vec n|^3} \right]^{\rm UV-div} +  \overbracket[0.8pt][0pt]{\sum_{\vec n\neq0}} \frac{1}{|\vec n|^3} 
\end{equation}
where 
\begin{equation}\label{sumtopo3}
        \overbracket[0.8pt][0pt]{\sum_{\vec n\neq0}} \frac{1}{|\vec n|^3}  \equiv \left[ \sum_{\vec n\neq0}\frac{1}{|\vec n|^3} \right]^{\rm tp}
\end{equation}
denotes the finite, scheme-independent topology-dependent contribution. For the purposes of the present work, its explicit value is not required, since it always appears together with the zero-mode contribution, which is itself left unfixed in our analysis. \\

Substituting the topology-dependent sums in Eqs. \eqref{eq:Sumsfinal} and \eqref{sumtopo3} into the energy density and pressure in Eq. \eqref{eq:ComponentsApp}, we obtain
\begin{equation}\label{Tmunu TP App}
    \begin{aligned}
        & \rho^{\mathrm{imp,tp}}_{\nu=\frac{3}{2}} =  - \frac{ \eta^4 H^4}{ 2\pi^2 L^4 }    \sum_{\vec r\neq \vec 0} \frac{1}{|\vec r|^4}  + \left( 1 - 18 \xi \right) \frac{\eta^2 H^4}{8 \pi  L^2}  \left[  \frac{1}{\pi} \sum_{\vec{r} \neq 0} \frac{e^{-\lambda L^2 |\vec{r}|^2}}{|\vec{r}|^2}+\sum_{\vec{n} \neq 0} \frac{\operatorname{erfc}\left(\frac{\pi \left|\vec{n}\right|}{L \sqrt{\lambda}}\right)}{|\vec{n}|} -\frac{2 \sqrt{\pi}}{L } \frac{1}{\sqrt{\lambda}}  - \frac{\lambda L^2}{\pi}  \right] \\
        &\quad - \frac{3 H^4\xi}{16 \pi^3 } \, \, \overbracket[0.8pt][0pt]{\sum_{\vec n\neq0}} \frac{1}{|\vec n|^3}  -\frac{3H\xi}{L^3}|b_{\vec 0}|^2 +(9-42\xi)\frac{H^7|\eta|^6}{18 L^3}|c_{\vec 0}|^2 +\frac{4H^4\xi\eta^3}{L^3} \left(   b_{\vec 0}^*c_{\vec 0}+b_{\vec 0}c_{\vec 0}^* \right)\,,\\
        & P^{\mathrm{imp,tp}}_{\nu=\frac{3}{2}}{}_i{}^j  =  - \frac{ \eta^4 H^4}{6\pi^2 L^4} \sum_{\vec r\neq \vec 0}  \frac{1}{|\vec r|^4}\delta_i{}^j \\
        & \quad - \left(1-18\xi\right) \frac{\eta^2 H^4}{24\pi L^2} \left[  \frac{1}{\pi} \sum_{\vec{r} \neq 0} \frac{e^{-\lambda L^2 |\vec{r}|^2}}{|\vec{r}|^2}+\sum_{\vec{n} \neq 0} \frac{\operatorname{erfc}\left(\frac{\pi \left|\vec{n}\right|}{L \sqrt{\lambda}}\right)}{|\vec{n}|} -\frac{2 \sqrt{\pi}}{L } \frac{1}{\sqrt{\lambda}}  - \frac{\lambda L^2}{\pi}  \right] \delta_i{}^j  + \frac{3H^4\xi}{16\pi^3} \, \, \overbracket[0.8pt][0pt]{\sum_{\vec n\neq0}} \frac{1}{|\vec n|^3}   \delta_i{}^j \\
        & \quad + \frac{3H\xi}{L^3}|b_{\vec 0}|^2 \delta_i{}^j+ (9-42\xi) \frac{H^7|\eta|^6}{18L^3} |c_{\vec 0}|^2 \delta_i{}^j\, .
    \end{aligned}
\end{equation}
Using the rescaling of the zero-mode integration constants in Eq. \eqref{rescaling} and identifying the scalar pressure according to Eq. \eqref{pressureRed}, Eq. \eqref{Tmunu TP App} reproduces the result in Eq. \eqref{Tmunu TP}.

\subsection{Ewald summation}\label{Ewald}

We now give the derivation of the Ewald representation \cite{Ewald} for the $|\vec{l}_r|^{-2}$ sum
\begin{equation}\label{Ewald1}
    \begin{aligned}
      & \sum_{\vec{r}\neq \vec{0} }  \frac{1}{|\vec{l}_r|^2}   = \sum_{\vec{r} \neq 0} \frac{e^{-\lambda\left|\vec{l}_r\right|^2}}{|\vec{l}_r|^2}+\frac{2 \pi^2}{\operatorname{Vol}} \sum_{\vec{n} \neq 0} \frac{1}{\left|k_n\right|} \operatorname{erfc}\left(\frac{\left|k_n\right|}{2 \sqrt{\lambda}}\right)-\frac{2 \pi^{3 / 2}}{\operatorname{Vol} } \frac{1}{\sqrt{\lambda}} +\frac{2 \pi^{3 / 2}}{L^2 }\lim_{\epsilon \to 0} \frac{1}{\sqrt{\epsilon}} - \lambda \,.
    \end{aligned}
\end{equation}
This example makes explicit how the artificial zero-mode divergence is isolated and how the finite part entering the main calculation is defined. \\

For the cubic torus, $|\vec l_{\vec r}|=L|\vec r|$. We introduce the dimensionless integration variable $u$ and the dimensionless Ewald splitting parameter $\mu = \lambda L^2$. We then write
\begin{equation}\label{Ewald2}
     \sum_{\vec{r}\neq \vec{0} }\frac{1}{|\vec{l}_r|^2}   = \sum_{\vec{r}\neq \vec{0} }\int_0^\infty \frac{du}{L^2} e^{-u |\vec{r}|^2} = \sum_{\vec{r}\neq \vec{0} } \frac{1}{L^2} \left[\int_0^\mu du e^{-u |\vec{r}|^2}+\int_\mu^\infty du e^{-u |\vec{r}|^2}\right]=\sum_{\vec{r}\neq \vec{0} }\int_0^\mu \frac{du}{L^2} e^{-u |\vec{r}|^2}+ \sum_{\vec{r}\neq \vec{0} }\frac{e^{-\mu\left|\vec{r}\right|^2}}{L^2|\vec{r}|^2}\,,
\end{equation}
where the first term in Eq. \eqref{Ewald2} contains the artificial IR divergence. We rewrite it as
\begin{equation}\label{Ewald3}
    \begin{aligned}
    \frac{1}{L^2}\sum_{\vec r\neq\vec 0}
    \int_0^\mu du\,e^{-u|\vec r|^2}  &=  \frac{1}{L^2}
    \left[   \sum_{\vec r}  \int_0^\mu du\,e^{-u|\vec r|^2} -  \int_0^\mu du \right] \\
    &=  \frac{1}{L^2} \left[  \int_0^\mu du  \sum_{\vec r}e^{-u|\vec r|^2}  -\mu  \right]\,.
    \end{aligned}
\end{equation}
Applying the Poisson summation formula gives
\begin{equation}\label{Ewald4}
    \sum_{\vec r\in\mathbb Z^3}e^{-u|\vec r|^2}  =
    \left(\frac{\pi}{u}\right)^{3/2}  \sum_{\vec n\in\mathbb Z^3}  e^{-\pi^2|\vec n|^2/u}  = \left(\frac{\pi}{u}\right)^{3/2}
    \sum_{\vec n\neq\vec 0}   e^{-\pi^2|\vec n|^2/u}
    +  \left(\frac{\pi}{u}\right)^{3/2} \, .
\end{equation}
Substituting Eq. \eqref{Ewald4} into Eq. \eqref{Ewald3}, we obtain
\begin{equation}\label{Ewald5}
    \begin{aligned}
    \frac{1}{L^2}\sum_{\vec r\neq\vec 0} \int_0^\mu du\,e^{-u|\vec r|^2}  &=  \frac{1}{L^2} \int_0^\mu du  \left(\frac{\pi}{u}\right)^{3/2}  \sum_{\vec n\neq\vec 0}  e^{-\pi^2|\vec n|^2/u} +\lim_{\epsilon \to 0}\frac{1}{L^2} \int_\epsilon^\mu du  \left(\frac{\pi}{u}\right)^{3/2}  -\frac{\mu}{L^2}
    \\
    &= \frac{\pi}{L^2}  \sum_{\vec n\neq\vec 0} \frac{1}{|\vec n|} \operatorname{erfc}  \left( \frac{\pi|\vec n|}{\sqrt{\mu}}   \right)  -  \frac{2\pi^{3/2}}{L^2\sqrt{\mu}} + \frac{2\pi^{3/2}}{L^2} \lim_{\epsilon \to 0}\frac{1}{\sqrt{\epsilon}} -   \frac{\mu}{L^2}\,.
    \end{aligned}
\end{equation}
The divergent term arises from the $\vec n=\vec 0$ contribution. Substituting Eq. \eqref{Ewald5} back into Eq. \eqref{Ewald2}, and using $\mu=\lambda L^2$, reproduces Eq. \eqref{Ewald1}, completing the derivation of the stated Ewald identity. \\

It is also useful to make explicit that the finite part of Eq. \eqref{Ewald1}, which corresponds to $B_1$ defined in Eq. \eqref{brakets}, is independent of the arbitrary Ewald splitting parameter. In terms of the dimensionless parameter $\mu=\lambda L^2$, we have
\begin{equation}
    B_1(\mu) = \frac{1}{\pi}\sum_{\vec r\neq0}
    \frac{e^{-\mu|\vec r|^2}}{|\vec r|^2} + \sum_{\vec n\neq0}
    \frac{\operatorname{erfc}\left(\frac{\pi|\vec n|}{\sqrt{\mu}}\right)}
    {|\vec n|} - \frac{2\sqrt{\pi}}{\sqrt{\mu}} - \frac{\mu}{\pi}\,.
\end{equation}
Differentiating term by term, we obtain
\begin{equation}
    \begin{aligned}
    \frac{dB_1}{d\mu}   = &-\frac{1}{\pi} \sum_{\vec r\in\mathbb Z^3}  e^{-\mu|\vec r|^2} + \frac{\sqrt{\pi}}{\mu^{3/2}}   \sum_{\vec n\in\mathbb Z^3}  e^{-\pi^2|\vec n|^2/\mu} \, ,
    \end{aligned}
\end{equation}
where the two subtraction terms have been combined with the zero modes of the corresponding Gaussian sums. The Poisson summation formula gives
\begin{equation}
    \sum_{\vec r\in\mathbb Z^3}  e^{-\mu|\vec r|^2} = \frac{\pi^{3/2}}{\mu^{3/2}}\sum_{\vec n\in\mathbb Z^3}   e^{-\pi^2|\vec n|^2/\mu} \, ,
\end{equation}
and therefore
\begin{equation}
    \frac{dB_1}{d\mu}=0 \, .
\end{equation}
Thus, the dependence of the separate real-space, reciprocal-space, and subtraction terms on the dimensionless Ewald splitting parameter $\mu$ cancels in their sum. 

\clearpage

\bibliographystyle{utphys}
\bibliography{main} 

\end{document}